\documentclass[11pt,a4paper]{article}
\usepackage[inactive, tightpage]{preview}
\usepackage{jheppub}
\usepackage{amsthm}
\usepackage[T1]{fontenc}
\usepackage[utf8]{inputenc}
\usepackage{mathtools}   
\usepackage{amsfonts}
\usepackage{physics}
\usepackage{mathrsfs}
\usepackage{natbib}
\usepackage{comment}
\usepackage{tikz, pgfplots}
\usepackage{ytableau}
\usetikzlibrary{patterns, calc, patterns, angles, quotes, decorations, shadows}
\usepackage{enumitem}
\usepackage{subcaption}
\usepackage[font={small}]{caption}
\usepackage{graphicx}
\usepackage{amssymb}
\usepackage{microtype}
\usepackage{todonotes}

\usepackage{microtype}
\usepackage{hyperref}

\definecolor{darkmidnightblue}{rgb}{0.0, 0.2, 0.4}
\hypersetup{
    colorlinks = true, 
    linkcolor=darkmidnightblue,
    citecolor=darkmidnightblue
}

\usepackage{cleveref}

\usepackage{booktabs}
\usepackage{pdflscape}
\usepackage{afterpage}
\usepackage{flafter}
\usepackage{rotating}
\usepackage{fancyhdr}
\usepackage{slashed}
\fancypagestyle{lscape}{%
\fancyhf{} 
\fancyfoot[LE]{}
\fancyfoot[LO] {}
 
}

\newcommand{\Z}{\mathbb{Z}}
\newcommand{\R}{\mathbb{R}} 
\newcommand{\C}{\mathbb{C}} 
\newcommand{\ch}{\text{ch}}
\DeclareMathSymbol{:}{\mathord}{operators}{"3A}

\title{Sinh Deformed Nakajima Operators}

\author[a]{Boan Zhao,}
\author[a]{Paul Luis Roehl,}
\author[b]{Chunhao Li}

\affiliation[a]{DAMTP, University of Cambridge, Cambridge, United Kingdom}

\affiliation[b]{Theoretical Physics Group, The Blackett Laboratory, Imperial College London,\\ Prince Consort Road
London, SW7 2AZ, UK}

\emailAdd{bz258@cam.ac.uk}
\emailAdd{plr30@cam.ac.uk}
\emailAdd{chunhao.li21@imperial.ac.uk}

\abstract{
We prove a novel action of the (three-dimensional) Heisenberg algebra on the equivariant K-theory of the Hilbert scheme of points on $\C^2$. These operators are defined via pushforwards and pullbacks via the Nakajima correspondences \cite{NakajimaMoreLectures} while tensoring the square roots of the canonical line bundles of the correspondences. We show, using supersymmetric localisation in 6d $(1, 1)$ Super Yang-Mills compactified on a circle,  that these operators correspond to instanton line operators wrapping the extra circle.
}

\theoremstyle{definition}

\begin{document}
\ytableausetup{boxsize = 0.2 cm}
\ytableausetup{centertableaux} 
\maketitle

\section{Introduction}
The geometry of instanton moduli spaces on $\C^2$ is a useful tool to construct representations of various algebras. An early example is due to Nakajima \cite{nakajima1999lectures, NakajimaMoreLectures} who constructed an action of the (infinite-dimensional) Heisenberg algebra on the equivariant cohomology of $U(1)$ instanton moduli space --- also known as the Hilbert scheme of points on $\C^2$. Later, several uplifts of Nakajima's operators to the equivariant K-theory of the same spaces have been found \cite{Schiffmann, NegutFlags, NegutElliptic}. These operators are used to prove the AGT correspondence \cite{GaiottoAGT}.

In this work, we prove a novel K-theoretic uplift of Nakajima operators. Our operators $a_\pm$ form a representation of the three-dimensional Heisenberg algebra $[a_-, a_+]\propto 1$, where $1$ stands for the identity operator. We briefly explain the mathematical definition of the operators in section \ref{sec: intro_def} and the physical realization as instanton line operators in 6d Super Yang-Mills in section \ref{sec: intro_phys}. 

\subsection{Mathematical Definition of the Operators $a_\pm$} \label{sec: intro_def}
The operators $a_\pm$ we will define rely on the geometry of $M_k$, the $U(1)$ $k$-instanton moduli space on $\C^2$ (also known as the Hilbert scheme of $k$-points on $\C^2$). A generic element $x_k \in M_k$ is a subset of $\C^2$ of size $k$ parametrizing the centres of the instanton. Our operators will be defined using the Nakajima correspondences (sometimes known as the nested Hilbert schemes) $M_{k, k + 1}\subset M_k \times M_{k + 1}$ constructed by adding the origin of $\C^2$. In other words, a generic element $(x_k, x_{k + 1})\in M_{k, k +1}$ satisfies $x_{k + 1} = x_k \cup \{(0, 0)\}$ where $x_k\in M_k, x_{k + 1}\in M_{k + 1}$ are subsets of $\C^2$ of sizes $k, k + 1$ respectively. $(0, 0)\in \C^2$ is the origin of $\C^2$.

The definition of $a_\pm $ also requires the following natural $U(1)^2$ action on $\C^2$:
\begin{eqnarray}\label{eq: torus_action}
    (z_1, z_2)\to (q_1z_1, q_2z_2)\quad (z_1, z_2)\in \C^2\quad (q_1, q_2)\in U(1)^2
\end{eqnarray}
which induces an action on all finite subsets of $\C^2$ and therefore induces an action on $M_k, k\geq 0$. The torus fixed points of $M_k$ are labelled by Young diagrams $Y_k$ with $k$ boxes. We use the convention that the subscripts label the number of boxes in a Young diagram. A torus fixed points of $M_{k, k + 1}$ is labelled by a pair of Young diagrams $(Y_k, Y_{k + 1})$ of sizes $k$ and $k+1$ respectively such that $Y_k\subset Y_{k + 1}$ (i.e. $Y_{k + 1}$ can be obtained from $Y_k$ by adding a box).

We define operators $a_\pm$ which act on an infinite dimensional vector space with a basis labelled by Young diagrams (torus fixed points of $M_k, k\geq 0$): $\ket{\emptyset}, \ket{\ydiagram{1}}, \ket{\ydiagram{2}}, \ket{\ydiagram{1, 1}},...$ as follows:
\begin{eqnarray}
    a_+ \ket{Y_k} = \sum_{Y_{k + 1}\supset Y_k}\frac{\hat{a}(TM_{k + 1}(Y_{k + 1}))}{\hat{a}(TM_{k, k + 1}(Y_k, Y_{k + 1}))}\ket{Y_{k + 1}}
\end{eqnarray}
where $TM_{k + 1}(Y_{k + 1})$ denotes the character (a Laurent polynomial in $q_1, q_2$ with integer coefficients) of the $U(1)^2$ acting on the tangent space of $M_{k + 1}$ at $Y_{k + 1}$. A similar meaning is attached to $TM_{k, k + 1}(Y_k, Y_{k + 1})$. The a-hat genus $\hat{a}$ of a Laurent polynomial in $q_1, q_2$ (with integer coefficients) is another function of $q_1, q_2$ defined by the following rules: $\hat{a}$ turns addition into multiplication (i.e. $\hat{a}(q_1 + q_2) =\hat{a}(q_1)\hat{a}(q_2) $) and satisfies
\begin{eqnarray}\label{eq: ahat_def}
    \hat{a}(nq_1^aq_2^b) = (q_1^{a/2}q_2^{b/2} - q_1^{-a/2}q_2^{-b/2})^n
\end{eqnarray}
for any integers $a,b, n$. For example,
\begin{eqnarray}
\begin{aligned}
    &\hat{a}(7q_1q_2^2 + 3q_1^{-2} - 5q_1q_2^3)\\
    = &\hat{a}(7q_1q_2^2)\hat{a}(3q_1^{-2})\hat{a}(-5q_1q_2^3)\\
    =&(q_1^{1/2}q_2 - q_1^{-1/2}q_2^{-1})^7(q_1^{-1} - q_1)^3(q_1^{1/2}q_2^{3/2} - q_1^{-1/2}q_2^{-3/2})^{-5}
\end{aligned}
\end{eqnarray}
Hence, all matrix coefficients of $a_+$ are functions of $q_1, q_2$ (not necessarily rational due to the possible half integer powers).
The $\hat{a}$-genus is (up to a factor of 2) a $\sinh$, hence we call our operators "sinh deformed Nakajima operators".  The definition of $a_-$ is
\begin{eqnarray}
    a_- \ket{Y_k} = \sum_{Y_{k - 1}\subset Y_k}\frac{\hat{a}(TM_{k - 1}(Y_{k - 1}))}{\hat{a}(TM_{k - 1, k}(Y_{k - 1}, Y_k))}\ket{Y_{k - 1}}
\end{eqnarray}
The first main result of this work is the following commutator:
\begin{eqnarray}\label{eq: commutator}
    [a_-, a_{+}] = (\sqrt{q_1} - \sqrt{q_1}^{-1})(\sqrt{q_2} - \sqrt{q_2}^{-1}) = \hat{a}(q_1 + q_2)
\end{eqnarray}

We can rewrite the operators $a_\pm$ in a more geometric way to allow a comparison with existing operators in the literature \cite{NegutElliptic}. Let \begin{eqnarray}
\begin{aligned}
    \pi_k:&\quad M_{k ,k + 1}\to M_k \\
    \pi_{k + 1}:&\quad M_{k ,k + 1}\to M_{k + 1}
    \end{aligned}
\end{eqnarray}
be the natural projection maps. The infinite-dimensional vector space described above is identified with the equivariant K-theory of $M_k$ (denoted $K_{q_1, q_2}(M_k)$) induced from the torus action \eqref{eq: torus_action}. The basis elements are identified with (suitably normalized \footnote{The normalization is that $\ket{Y}$ pullbacks to 1 under the inclusion map of the fixed point $Y$.}) skyscraper sheaves of the fixed points of $M_k, k\geq 0$. The operators
$a_\pm: K_{q_1, q_2}(M_k)\to K_{q_1, q_2}(M_{k \pm  1}$) are defined by:
\begin{eqnarray}
\begin{aligned}
    a_+ = \sqrt{KM_{k + 1}}^{-1}\otimes (\pi_{k + 1})_*(\sqrt{KM_{k, k + 1}}\otimes \pi_k^*)\\
    a_- = \sqrt{KM_{k - 1}}^{-1}\otimes (\pi_{k - 1})_*(\sqrt{KM_{k - 1, k}}\otimes \pi_{k}^*)
\end{aligned}
\end{eqnarray}
where $\sqrt{KM_{k, k +1}}, \sqrt{KM_{k}}$ are the square roots of the canonical line bundle of $M_{k ,k + 1}$ and $M_k$ respectively. The standard operators in the K-theoretic uplifts \cite{NegutElliptic} are simply $(\pi_{k\pm  1})_*\pi_k^*$ (possibly tensoring with an \textbf{integer} power of $KM_{k, k + 1}, KM_{k- 1, k}$). It is known that $\sqrt{KM_{k\pm 1}}^{-1} = \sqrt{q_1q_2}^{k\pm 1}$ which is merely a normalization factor for $a_\pm$ at fixed $k$. Therefore, the novel part of our definition is the tensor product with the square root (a \textbf{half-integer} power
\footnote{
    Another way of writing the operators is as follows. The canonical line bundle $KM_{k ,k + 1}$ equals $q_1^{-k}q_2^{-k}L_{k, k +1}$ as a K-theory class, where the tautological line bundle $L_{k, k + 1}\to M_{k, k + 1}$ has weight equal to $Y_{k + 1} - Y_k$ at the fixed point $(Y_k, Y_{k + 1})$. As a result we have
    \begin{eqnarray}
        \begin{aligned}
            a_+  &= \sqrt{q_1q_2}(\pi_{k + 1})_* (\sqrt{L_{k,   k + 1}}\otimes \pi_k^*)\\
            a_-  &= (\pi_{k - 1})_* (\sqrt{L_{k - 1, k }}\otimes \pi_{k }^*)
        \end{aligned}
    \end{eqnarray}
}
) of the canonical line bundle of the correspondence before pushing forward via $\pi_{k \pm 1}$. Despite the fact that these operators are not linear combinations of known operators, we will prove that their commutator can be reduced to a known commutator in a simple way. 

\subsection{$a_\pm$ as Instanton Operators in 6d Super Yang-Mills} \label{sec: intro_phys}

Our next result is an argument that $a_\pm$ naturally arise in 6d $(1, 1)$ Super Yang-Mills on $\C^2 \times [0, 1] \times S^1$ with a twisted periodicity along the $S^1$. We think of $[0, 1]$ as time and $\C^2 \times S^1$ as space. The low energy dynamics of the theory is given by maps from $S^1\times [0, 1]\to M_k, k\geq 0$ (i.e. strings propagating on instanton moduli spaces), where the instantons live on the $\C^2$ factor. The BPS Hilbert space naturally has a basis labelled by Young diagrams/torus fixed points of $M_k$ which correspond to strings localized at these points. The operators $a_\pm$ are identified with instanton line operators $a_\pm^{6d}$ wrapping the $S^1$ and acting on the BPS Hilbert space. These line operators are located at a chosen point in $\C^2\times [0, 1]$ and fill up the entire $S^1$. They are defined by the property that the instanton number on a four-sphere surrounding the line operator is $\pm 1$.

We compute the matrix elements of these instanton line operators $a_\pm^{6d}$ using supersymmetric localisation in the 6d gauge theory. We prepare a boundary state $\ket{Y_k}$ on the first boundary $\C^2 \times 0 \times S^1$ and another boundary state $\ket{Y_{k + 1}}$ on the second boundary $\C^2 \times 1 \times S^1$. The supersymmetric path integral with the instanton line operator $a_+^{6d}$ insertion computes the matrix element
\begin{eqnarray}\label{eq: matrix_elem}
    \bra{Y_{k + 1}} a_{+}^{6d} \ket{Y_k} = \frac{1}{\hat{a}(TM_{k, k + 1}(Y_k, Y_{k + 1}))}
\end{eqnarray}
The denominator is the $\hat{a}$ \eqref{eq: ahat_def} of the tangent character of $M_{k ,k + 1}$ at the torus fixed point $(Y_k, Y_{k + 1})$. A similar result holds for $a_-$. One can compute the norm square $\braket{Y_k}{Y_k}$ by computing the supersymmetric path integral with $\ket{Y_k}$ at both boundaries and \textbf{no} instanton line operator insertion. The result is
\begin{eqnarray}\label{eq: overlap}
    \braket{Y_k}{Y_k} = \frac{1}{\hat{a}(TM_k(Y_k))}
\end{eqnarray}
Combining the two equations, we obtain $a_+ = a_+^{6d}$. And a similar argument shows $a_- = a_-^{6d}$.

The localisation computation can also be performed directly in the infrared dynamics of the 6d theory --- a 2d $\sigma$-model into the instanton moduli spaces $M_k$. We place the 2d theory on a cylinder $[0, 1]\times S^1$ and insert a topological interface $a_+^{\text{2d}}$ at $0.5 \times S^1$. More precisely, we let the boson take values in $M_k$ when $x^0 \in [0, 0.5]$ and in $M_{k + 1}$ when $x^0 \in [0.5, 1]$, where $x^0$ is the time coordinate on $[0, 1]$. At time $x^0 = 0.5$ the boson jumps from $M_k$ to $M_{k + 1}$ via the correspondence $M_{k, k + 1}$. We will perform supersymmetric localisation with and without the topological interface and recover \eqref{eq: matrix_elem} \eqref{eq: overlap} (with $a_+^{6d}$ replaced by the 2d topological interface $a_+^{\text{2d}}$ regarded as a BPS operator). Hence we have $a_\pm = a_\pm^{6d}= a_\pm^{2d}$. We also perform localisation computations for more than two insertions of the topological interface and show that it correctly reproduces the composition of the operators $a_\pm^{2d} \circ a_\pm^{2d}$ even when the correspondences do not intersect transversely.

\section{Proof of the Commutator $[a_+, a_-]$}\label{ch: proof}
In this chapter, we define the main objects of study: the operators $a_\pm$ acting on a vector space with a basis labelled by Young diagrams. The operators are defined via the Nakajima correspondences $M_{k, k+1}\subset M_k \times M_{k + 1}$, where $M_k$ is the $U(1)$ $k$-instanton moduli space on $\C^2$. After a brief review of $M_k$ and $M_{k, k +1}$, we present the definitions of $a_\pm$ via its matrix elements \eqref{eq: aplus_def} \eqref{eq: aminus_def} on the basis described above. We then simplify the matrix elements and obtain \eqref{eq: aplusminus_def}. Finally, we prove the commutator $[a_- ,  a_+] = (q_1^{1/2} - q_1^{-1/2})(q_2^{1/2} - q_2^{-1/2})$ directly in the last two sections.

\subsection{Review of $U(1)$ Instanton Moduli Spaces}\label{sec: review_inst_moduli}
In this and the next section, we review the basic properties of $U(1)$ instanton moduli spaces $M_k$ on $\C^2$ (also known as the Hilbert scheme of points on $\C^2$) and the Nakajima correspondence $M_{k, k +1}\subset M_k \times M_{k + 1}$ (sometimes known as the nested Hilbert schemes). Readers familiar with the material can look at the definitions below and proceed to section \ref{sec: aplusminus_def} for the definition of the operators $a_\pm$. Readers interested in the proof of the commutator $[a_-, a_+]$ only can proceed directly to section \ref{sec: off_diagonal}.  We first review $M_k$ \cite{nakajima1999lectures} which can be defined via the ADHM construction:
\begin{eqnarray}
    M_k \cong \left.\left(\quad
\begin{aligned}
    &[B_1, B_2] + I J = 0\\
    &\text{span}((B_1)^n(B_2)^mI,m,n\geq 0) = \C^k
\end{aligned}
\quad\right)\middle/ GL(k;\C)\right.
\end{eqnarray}
where $B_1, B_2$ are complex $k\times k$ matrices transforming in the adjoint of $GL(k;\C)$. $I, J$ are complex column and row vectors of size $k$ transforming in the fundamental and antifundamental of $GL(k;\C)$. We also require that the span of $B_1^nB_2^mI$ for $m,n \geq 0$ is the whole $\C^k$. This condition guarantees the smoothness of the quotient.

The geometric picture behind this is that a generic element $x_k\in M_k$ is a subset of $\C^2$ of size $k$. There is no constraint on the $k$ elements and hence the (complex) dimension of $M_k$ is $2k$.
We write a generic element $x_k\in M_k$ as $x_k = \{p_1,..., p_k\}$ where $p_i\in \C^2$ and $ \{p_1,..., p_k\} =  \{p_{\sigma(1)},..., p_{\sigma(k)}\}$ in $M_k$ for any permutation $\sigma$ in the symmetric group of $k$ elements. For example, $\{(3, -\sqrt{3}), (-10, \pi), (1,2)\} =  \{(1,2), (3, -\sqrt{3}), (-10, \pi)\}$ in $M_3$. Generically, the joint eigenvalues of $B_1, B_2$ parametrize the locations of the $k$ points in $x_k$ and $I$ can be chosen to be $(1,1,...,1)$. One can prove that $J = 0$ \cite{nakajima1999lectures}. For example, the point $\{(1,2), (3, -\sqrt{3}), (-10, \pi)\}$ corresponds to \footnote{The choice of the matrix data is not unique as there is a residual $S_3$ action of $GL(3; \C)$ which permutes the diagonal elements of $B_1, B_2$ simultaneously.}
\begin{eqnarray}
    B_1 = 
    \begin{pmatrix}
        1 & 0 & 0 \\
        0 & 3 & 0 \\
        0 & 0 & -10
    \end{pmatrix}\quad
    B_2 = 
    \begin{pmatrix}
        2 & 0 & 0 \\
        0 &-\sqrt{3} & 0 \\
        0 & 0 & \pi
    \end{pmatrix}
    \quad
    I = 
    \begin{pmatrix}
        1\\
        1\\
        1
    \end{pmatrix}
    \quad J = 0
\end{eqnarray}
There are special points \footnote{To parametrize all points in $M_k$ we should think of $x_k$ as ideals in the polynomial ring $\C[z_1, z_2]$ of codimension (as a complex vector space) $k$.} in $M_k$ which correspond to the case when two or more of the $p_i$ in $x_k$ collide. $M_k$ remains smooth in that case via a generalization of the blowup operation. $B_1, B_2$ are not necessarily diagonalisable at these points and one often needs to put them into Jordan normal forms.

The torus action \eqref{eq: torus_action} induces an action on the matrix data:
\begin{eqnarray}
    B_1\to q_1B_1 \quad B_2\to q_2B_2 \quad I\to I \quad J\to q_1q_2J
\end{eqnarray}
The torus fixed points are labelled by Young diagrams $Y_k$ in the following way: $\C^k$ has a basis labelled by boxes of the Young diagrams $\ket{p, q}$ where $(p, q)$ denotes the coordinates of the box. $I  = \ket{0,0}$, the box at the top left. The actions of $B_1, B_2$ are given by:
\begin{eqnarray}
    B_1\ket{p, q} = \ket{p + 1, q} \quad B_2\ket{p, q} = \ket{p, q + 1}
\end{eqnarray}
If $\ket{p, q}$ lies outside the Young diagram, it is treated as zero. Here is an example of a Young diagram with the coordinates $(p, q)$ of all the boxes labelled:
\begin{eqnarray*}
  \ytableausetup{boxsize=2.5em}
  Y =
  \begin{ytableau}
    (0, 0) & (1, 0) & (2, 0) \\
    (0, 1)
  \end{ytableau}
\end{eqnarray*}
The actions of $B_1, B_2$ are:
\begin{eqnarray}
\begin{aligned}
    &B_1\ket{0, 0} = \ket{1, 0}\quad B_1\ket{1, 0} = \ket{2, 0}\quad B_1\ket{2, 0} = B_1\ket{0, 1} = 0\\
    &B_2\ket{0, 0} = \ket{0, 1}\quad B_2\ket{1, 0} = B_2\ket{2, 0} = B_2\ket{0, 1} = 0
\end{aligned}
\end{eqnarray}
The proof of the Young tableau representation follows from the equation
\begin{eqnarray}\label{eq: def_g}
    q_1B_1 = gB_1g^{-1} \quad q_2B_2 = gB_2g^{-1}\quad gI = I\quad Jg^{-1} =q_1q_2J
\end{eqnarray}
for some $g = g(q_1, q_2)$ when $(B_1, B_2, I, J)$ is a torus fixed point. The states $\ket{p, q}$ are the weight vectors of $g(q_1, q_2)$:
\begin{eqnarray}\label{eq: g_basis}
    g(q_1, q_2) \ket{p, q} = q_1^pq_2^q\ket{p, q}
\end{eqnarray}
and the actions of $B_1, B_2$ on the weight vectors can be inferred from \eqref{eq: def_g}.
\subsection{Review of Nakajima Correspondences}\label{sec: review_corresp}
Next we move onto the Nakajima correspondences (sometimes known as nested Hilbert schemes). An informal definition
\footnote{A rigorous definition of the correspondence is as follows:
\begin{eqnarray}
    M_{k, k + 1} =
    \{x_{k + 1}\subset x_k\quad 
    \text{supp}(x_k / x_{k + 1})\subset (0,0)\in \C^2
    \quad x_{k + 1}\in M_{k+ 1}\quad x_k\in M_k\}
\end{eqnarray}
where we think of $x_k, x_{k + 1}$ as ideals of codimension (as a complex vector space) $k, k +1$ in the polynomial ring $\C[z_1, z_2]$. We identify $x_k, x_{k + 1}$ with the corresponding ideal sheaves on $\C^2$ and $\text{supp}$ is the support of the quotient of two ideal sheaves (the locus where the quotient is nonzero).
} of $M_{k, k + 1}\subset M_k \times M_{k + 1}$ is as follows:
\begin{eqnarray}
    M_{k, k + 1} \approx \{(x_k, x_{k + 1}) \quad x_{k + 1} = x_k \cup \{(0,0)\} \quad x_k\in M_k \quad x_{k + 1}\in M_{k + 1}\}
\end{eqnarray}
Hence $M_{k, k + 1}$ generically consists of all pairs $(x_k, x_{k + 1})$ such that $x_{k + 1}$ is the union of $x_k$ with the origin of $\C^2$. As an example, take $x_2 = \{(2, -3), (1, \sqrt{5})\}\in M_2$ and $x_3 = \{(2, -3), (1, \sqrt{5}), (0, 0)\}\in M_3$, then the pair $(x_2, x_3)$ lies in $M_{2, 3}$. Generically, $M_{k , k + 1}$ is uniquely determined by $x_k$ hence its dimension is $2k$. Subtleties arise again when one of the points in $x_k$ collide with the origin. $M_{k , k+ 1}$ remains smooth in this process.

The ADHM construction for $M_{k , k+ 1}$ was presented in \cite{NegutFlags} and will play a crucial role in computing the matrix elements of $a_\pm$. The matrix data consists of
\begin{eqnarray}
    M_{k , k+ 1} \cong \left.\left(\quad
\begin{aligned}
    &[B_1', B_2'] + I' J' = 0\\
    &\text{span}((B_1')^n(B_2')^mI',m,n\geq 0) = \C^{k + 1}
\end{aligned}
\quad\right)\middle/ G(k, k + 1)\right.
\end{eqnarray}
where the matrix data and the group $G(k, k+ 1)$ have the following form:
\begin{eqnarray}
\begin{aligned}
    &B_1' = \begin{pmatrix}
    0 & * & \cdots & * \\
    0 & * & \cdots & * \\
    \vdots & \vdots & \ddots & \vdots \\
    0 & * & \cdots & *
\end{pmatrix}
\quad
B_2' = \begin{pmatrix}
    0 & * & \cdots & * \\
    0 & * & \cdots & * \\
    \vdots & \vdots & \ddots & \vdots \\
    0 & * & \cdots & *
\end{pmatrix}
\quad I' = \begin{pmatrix} * \\ * \\ \vdots \\ * \end{pmatrix} \quad
J' = \begin{pmatrix} 0 & * & \cdots & * \end{pmatrix}\\
&G(k, k + 1) = 
\begin{pmatrix}
    * & * & \cdots & * \\
    0 & * & \cdots & * \\
    \vdots & \vdots & \ddots & \vdots \\
    0 & * & \cdots & *
\end{pmatrix}
\end{aligned}
\end{eqnarray}
where $*$ denotes an element which is not required to vanish.
We can recover $x_k$ from the matrix data as follows:
\begin{eqnarray}
\begin{aligned}
    &B_1' = \left(
  \begin{array}{c|ccc}
    0 & * & \cdots & *  \\
    \hline
    0 & \\
    \vdots & & B_1(x_{k})\\
    0 & 
  \end{array}
\right)\quad
B_2' = \left(
  \begin{array}{c|ccc}
    0 & * & \cdots & *  \\
    \hline
    0 & \\
    \vdots & & B_2(x_{k })\\
    0 & 
  \end{array}
\right)\\
 &I' = \left(
  \begin{array}{c}
    *\\
    \hline
    I(x_k)
  \end{array}
\right)\quad
J' = (0 | J(x_k))
\end{aligned}
\end{eqnarray}
 In other words, the matrix data $B_1(x_k), B_2(x_k), I(x_k), J(x_k)$ associated with $x_k$ is constructed by removing the first rows and columns of $B_1', B_2'$ and the first elements of $I'$ and $J'$ respectively.
The matrix data $B_1(x_{k + 1}), B_2(x_{k + 1}), I(x_{k + 1}), J(x_{k + 1})$ for $x_{k + 1}$ are the same as $B_1', B_2', I', J'$ respectively. However, one should note that the map $M_{k , k + 1}\to M_{k + 1}$ is not injective as the ADHM data for the target space is quotiented by a larger group $GL(k + 1;\C)$.

Here is an example of the ADHM construction for $M_{k , k + 1}$. Take $x_2 = \{(2, -3), (1, \sqrt{5})\}\in M_2$ and $x_3 = \{(2, -3), (1, \sqrt{5}), (0, 0)\}\in M_3$, we know that $(x_2, x_3)\in M_{2, 3}$. The corresponding matrix data are:
\begin{eqnarray}
   B_1' = 
    \begin{pmatrix}
        0 & 0 & 0 \\
        0 & 2 & 0 \\
        0 & 0 & 1
    \end{pmatrix}\quad
    B_2' = 
    \begin{pmatrix}
        0 & 0 & 0 \\
        0 & -3 & 0 \\
        0 & 0 & \sqrt{5}
    \end{pmatrix}
    \quad
    I' = 
    \begin{pmatrix}
        1\\
        1\\
        1
    \end{pmatrix}
    \quad J' = 0
\end{eqnarray}

A point $(x_k, x_{k + 1})$ of $M_{k, k + 1}$ is fixed by the torus action if and only if both $x_k$ and $x_{k + 1}$ are fixed by the torus action. Therefore, torus fixed points of $M_{k, k + 1}$ are labelled by pairs of Young diagrams $(Y_k, Y_{k + 1})$. It can be shown \cite{NakajimaMoreLectures} that we need the additional constraint $Y_k\subset Y_{k + 1}$.

\subsection{Matrix Elements of $a_\pm$}\label{sec: matrix_elem}\label{sec: aplusminus_def}
In this section, we define the main objects of interest: the operators $a_\pm$ acting on a complex vector space with a basis parametrized by torus fixed points of $M_k,k \geq 0$ (Young diagrams): $\ytableausetup{boxsize=0.8em}\ket{\emptyset}, \ket{\ydiagram{1}}, \ket{\ydiagram{2}},\ket{\ydiagram{1, 1}},...$. The operators $a_+$ will be defined via their matrix elements on this basis as follows and $a_-$ will be defined at the end of this section:
\begin{eqnarray}\label{eq: aplus_def}
    a_+ \ket{Y_k} = \sum_{Y_{k + 1}\supset Y_k}\frac{\hat{a}(TM_{k + 1}(Y_{k + 1}))}{\hat{a}(TM_{k, k + 1}(Y_k, Y_{k + 1}))}\ket{Y_{k + 1}}
\end{eqnarray}
where $Y_k, Y_{k + 1}$ are Young diagrams of size $k, k + 1$ respectively and we sum over all $Y_{ k+ 1}$ which contains $Y_k$. $TM_k(Y_k)$ denotes the character of the torus action on the tangent space of $M_k$ at the fixed point $Y_k$. A similar meaning is attached to $TM_{k, k + 1}(Y_k, Y_{k + 1})$. For example, the tangent character at $(0, 0)\in M_1 = \C^2$ is $q_1 + q_2$. The definition of the $\hat{a}$ genus is \eqref{eq: ahat_def}. Since the number of boxes in a Young diagram is the instanton number $k$, we say that $a_+$ raises the instanton number by one \footnote{Similarly, $a_-$ defined later lowers the instanton number by one.}. Now we give an example of the matrix elements of $a_+$. More examples will appear at the end of this section.
\begin{eqnarray}
    a_+\ket{\emptyset} = \frac{
        \hat{a}(TM_1(\ydiagram{1}))
    }{
        \hat{a}(TM_{0, 1}(\emptyset, \ydiagram{1})
    }
    \ket{\ydiagram{1}} = \hat{a}(q_1 + q_2)\ket{\ydiagram{1}} = (q_1^{1/2} - q_1^{-1/2})(q_2^{1/2} - q_2^{-1/2})\ket{\ydiagram{1}}
\end{eqnarray}
The only Young diagram with one box which contains the empty diagram $\emptyset$ is $\ydiagram{1}$. The denominator $\hat{a}(TM_{0, 1}( \emptyset, \ydiagram{1})$ is 1 because $M_{0, 1}$ is a point which contains $(x_0 = \emptyset, x_1 = \{(0, 0)\})$, so its tangent character $TM_{0, 1}$ is zero. The a-hat genus of zero is one. The Young diagram $\ydiagram{1}$ correspond to the fixed point $(0, 0)\in M_1$ and we have the desired result for the numerator.

Now we derive a general formula for the matrix elements of $a_+$. To do this, we need to know the difference between the tangent characters $TM_{k + 1}(Y_{k + 1})$ and $TM_{k, k + 1}(Y_k, Y_{k + 1})$. To compute the tangent character $TM_{k + 1}$, we consider an infinitesimal fluctuation of the form
\begin{eqnarray}
    B_1\to B_1 + \delta B_1\quad B_2 \to B_2 + \delta B_2 \quad I \to I + \delta I \quad J \to J + \delta J
\end{eqnarray}
Under the torus action,
\begin{eqnarray}
    B_1 +  \delta B_1 \to q_1(B_1 + \delta B_1) = B_1 + q_1 g^{-1}\delta B_1 g 
\end{eqnarray}
where the equality holds in $M_{k + 1}$ due to the $g\in GL(k + 1;\C)$ action, where $g$ is defined in \eqref{eq: def_g} and is diagonal in the basis $\ket{p, q}$ with eigenvalues $q_1^pq_2^q$. Hence the contribution from $\delta B_1$ to the tangent character is $q_1 \tr(g)\tr(g^{-1})$. As an example, take the fixed point $\ydiagram{2}$. The eigenvalues of the matrix $g$ are $1, q_1$ which are the weights of the two boxes in $\ydiagram{2}$ \footnote{see the appendix for how we identify boxes with monomials in $q_1, q_2$.}. Hence $\delta B_1$ contributes $q_1(1 + q_1)(1 + q_1^{-1})$ to $TM_2(\ydiagram{2})$. Similarly, one can show that $\delta B_2, \delta I, \delta J$ contributes $q_2\tr(g)\tr(g^{-1}), \tr(g^{-1}), q_1q_2\tr(g)$ to the tangent characters. We also need to impose the ADHM equation $[B_1, B_2] + IJ = 0$ whose linearized version is
\begin{eqnarray}
    [B_1, \delta B_2] + [\delta B_1, B_2] + I \delta J + \delta I J = 0
\end{eqnarray}
This constraint transforms in the adjoint of $GL(k;\C)$ so contributes $-q_1q_2\tr(g)\tr(g^{-1})$ to the tangent character. Finally, the pure gauge transformation contributes $-\tr(g)\tr(g^{-1})$. The tangent character is:
\begin{eqnarray}
    TM_{k + 1}(Y_{k + 1}) = \tr(g)\tr(g^{-1})(q_1 + q_2 - q_1q_2 - 1) + \tr(g^{-1}) + q_1q_2\tr(g)
\end{eqnarray}
For example, to compute $TM_2(\ydiagram{2})$, we replace $\tr(g) = 1 + q_1, \tr(g^{-1}) = 1 + q_1^{-1}$ in the formula and obtain:
\begin{eqnarray}
    TM_2(\ydiagram{2}) = q_1 + q_2 + \frac{q_2}{q_1} + q_1^2
\end{eqnarray}

A similar story holds for $TM_{k, k + 1}(Y_k, Y_{k + 1})$. Let $g'$ be the corresponding diagonal matrix (i.e. $q_1B_1' = g'B_1g'^{-1}...$). $\delta B_1'$ does not contribute the full adjoint representation of $GL(k;\C)$. Instead, only the nonvanishing components of $\delta B_1'$ contribute. For example, take $k = 2$ and let $(Y_2, Y_3)\in M_{2,3}$ be a fixed point. The two matrices $g', \delta B_1'$ have the forms in the eigenbasis $\ket{p, q}$ \eqref{eq: g_basis} of $g'$:
\begin{eqnarray}
    g' = 
    \begin{pmatrix}
    g_1' & 0 & 0\\
    0 & g_2' & 0\\
    0 & 0 & g_3'
    \end{pmatrix}
    \quad
    \delta B_1' = 
    \begin{pmatrix}
    0 & * & *\\
    0 & * & *\\
    0 & * & *
    \end{pmatrix}
\end{eqnarray}
where $g_1'$ is the weight of the extra box $Y_3 - Y_2$ and 
$g_1',g_2'$ are the weights in $Y_2$.\footnote{This is true more generally. The top left components of $g'$ is the weight of the extra box $Y_{k + 1} - Y_k$ and the other diagonal components are the weights in $Y_k$.} The 6 starred components of $B_1$ have the following weights from the \textbf{inverse} adjoint action $\delta B_1'\to g'^{-1}\delta B_1' g'$:
\begin{eqnarray}
    \text{weight}(\delta B_1') = \begin{pmatrix}
    0 & g_2'/g_1' & g_3'/g_1'\\
    0 & 1 & g_3'/g_2'\\
    0 & g_2'/g_3' & 1
    \end{pmatrix}
\end{eqnarray}
Hence the contribution of $q_1g'^{-1}\delta B_1'g'$ to the tangent character is
\begin{eqnarray}
    q_1(1 + 1 + g_2'/g_1' + g_3'/g_1' + g_3'/g_2' + g_2'/g_3')
\end{eqnarray}

Our goal is to compute the difference between the tangent characters $TM_{k + 1}(Y_{k + 1}) - TM_{k, k + 1}(Y_k, Y_{k + 1})$. We do it term by term and first compute the difference in the contribution from $\delta B_1$ (from the tangent space of $Y_{k + 1}$) and $\delta B_1'$ (from the tangent space of $(Y_k, Y_{k + 1})$). The difference comes from the first column of $\delta B_1'$ and equals
\begin{eqnarray}
    q_1 \sum_{i\geq 1}\frac{g_1'}{g_i'} =  \sum_{s\in Y_{k + 1}}\frac{Y_{k + 1} - Y_k}{s}q_1  
\end{eqnarray}
For example, take $Y_3 = \ydiagram{2, 1}, Y_2 = \ydiagram{2}$. We have $Y_3 - Y_2 = q_2$ and $s$ sums over $1, q_1, q_2$.

One can also compute the difference between the contributions from $\delta B_2, \delta B_2'$ and other terms. The final result is
\begin{eqnarray}
\begin{aligned}
&TM_{k + 1}(Y_{k + 1}) - TM_{k, k + 1}(Y_k, Y_{k + 1})\\ 
=&
    q_1q_2 (Y_{k + 1} - Y_k) + 1 + \sum_{s\in Y_{k + 1}}\frac{Y_{k + 1} - Y_k}{s}(q_1 + q_2 - q_1q_2 - 1)
\end{aligned}
\end{eqnarray}
Now we substitute this expression into the definition of $a_+$ \eqref{eq: aplus_def} and obtain:
\begin{eqnarray}\label{eq: aplus_def_temp}
    a_+ \ket{Y} = \sum_{u\in \text{Add}(Y)}\hat{a}\left(q_1q_2 u + 1 + \sum_{s\in Y + u} \frac{u}{s}(q_1 + q_2 -q_1q_2 - 1) \right)\ket{Y + u}
\end{eqnarray}
where $u$ sums over addable boxes to $Y$ (see appendix \ref{app: young_diag}).
To make sense of this $\hat{a}$-genus, one needs to replace $u$ and $s$ by monomials in $q_1, q_2$ as in the appendix.

We now define $a_-$:
\begin{eqnarray}\label{eq: aminus_def}
    a_- \ket{Y_k} = \sum_{Y_{k - 1}\subset Y_k}\frac{\hat{a}(TM_{k - 1}(Y_{k - 1}))}{\hat{a}(TM_{k - 1, k}(Y_{k - 1}, Y_k))}\ket{Y_{k - 1}}
\end{eqnarray}
A similar computation shows that
\begin{eqnarray}
    a_-\ket{Y} =\sum_{u\in \text{Rm}(Y)}\hat{a}\left(-u^{-1} +1 - \sum_{s \in Y - u} \frac{s}{u}(q_1 + q_2 - q_1q_2 - 1)\right)\ket{Y - u}
\end{eqnarray}
where $u$ sums over removable boxes of $Y$ (see appendix \ref{app: young_diag}).

We now give some examples before moving on to the proof of their commutator. For $a_+$,
\begin{eqnarray}
\begin{aligned}
    &a_+\ket{\emptyset} = \sqrt{q_1q_2}(1 - q_1^{-1})(1 - q_2^{-1})\ket{\ydiagram{1}}\\
    &a_+\ket{\ydiagram{1}} = \sqrt{q_1}q_2(1-q_1^{-1})(1 - q_2^{-2})\ket{\ydiagram{1, 1}} + \sqrt{q_2}q_1(1-q_2^{-1})(1 - q_1^{-2})\ket{\ydiagram{2}} 
\end{aligned}
\end{eqnarray}
We derive the second equation as an example. $u$ sums over $q_1, q_2$ in \eqref{eq: aplus_def_temp} which corresponds to $\ket{\ydiagram{2}}, \ket{\ydiagram{1, 1}}$ respectively. The component on $\ket{\ydiagram{2}}$ is computed by setting $u = q_1$ in \eqref{eq: aplus_def_temp} and summing over $s = 1, q_1$ in the \eqref{eq: aplus_def_temp}. The result is $\hat{a}(q_1 + q_2^2)$ and $\hat{a}$ is defined in \eqref{eq: ahat_def}. Some examples for $a_-$ are
\begin{eqnarray}
\begin{aligned}
    a_-\ket{\emptyset} &= 0\\
    a_-\ket{\ydiagram{1}} &= \ket{\emptyset}\\
    a_-\ket{\ydiagram{1, 1}}& = \sqrt{q_2}(1 - q_1^{-1})(1 - q_2/q_1)^{-1}\ket{\ydiagram{1}}\\
    a_-\ket{\ydiagram{2}} &= \sqrt{q_1}(1 - q_2^{-1})(1 - q_1/q_2)^{-1}\ket{\ydiagram{1}}
\end{aligned}
\end{eqnarray}
One can check the equality \eqref{eq: commutator} to the first few basis elements using these formulas.

\subsection{Vanishing of Off-diagonal Components of $[a_-, a_+]$}\label{sec: off_diagonal}
We begin the proof of the main result \eqref{eq: commutator}. First we recall the setup. We have an infinite dimensional complex vector space with a basis labelled by Young diagrams: $\ytableausetup{boxsize=0.8em}\ket{\emptyset}$, $\ket{\ydiagram{1}}$, $\ket{\ydiagram{2}}$, $\ket{\ydiagram{1, 1}}$, .... We define operators $a_\pm$ acting on this vector space by their matrix elements in this basis:
\begin{eqnarray}\label{eq: aplusminus_def}
\begin{aligned}
    a_+ \ket{Y} = \sum_{u\in \text{Add}(Y)}\hat{a}\left(q_1q_2 u + 1 + \sum_{s\in Y + u} \frac{u}{s}(q_1 + q_2 -q_1q_2 - 1) \right)\ket{Y + u}\\
    a_-\ket{Y} =\sum_{u\in \text{Rm}(Y)}\hat{a}\left(-u^{-1} +1 - \sum_{s \in Y - u} \frac{s}{u}(q_1 + q_2 - q_1q_2 - 1)\right)\ket{Y - u}
\end{aligned}
\end{eqnarray}
$Y$ is a Young diagram. In the formula for $a_+$, $u$ sums over boxes which can be added to $Y$ such that $Y + u$ is still a Young diagram. In the formula for $a_-$, $u$ sums over boxes of $Y$ such that $Y - u$ is still a Young diagram (see appendix \ref{app: young_diag}). $s$ sums over boxes of $Y\pm u$. The reader can refer to section \eqref{sec: matrix_elem} for examples of these matrix elements.

In this section, we prove that $[a_-, a_+]$ is diagonal in the basis $\ket{Y}$. The proof is essentially an adaptation of proposition 18.1.2 of \cite{OkounkovQGQH} to K-theory and similar to the argument in \cite{FeiginShuffle, NegutFlags}. In other words, $[a_-, a_+]\ket{Y}\propto \ket{Y}$ for any Young diagram $Y$. Take two \textbf{distinct} Young diagrams $Y_k\neq  Y'_k$ with $k$ boxes each, where $k\geq  2$ \footnote{the action of $[a_-, a_+]$ on $\ket{\emptyset}, \ket{\ydiagram{1}}$ is clearly proportional to $\ket{\emptyset}, \ket{\ydiagram{1}}$ respectively.}. We will show that $[a_-, a_+]\ket{Y_k}$ has no  component in $\ket{Y'_k}$. The key idea is that we only need to consider the component of $a_+\ket{Y_k}$ on $\ket{Y_{k + 1}}$ and the component of $a_-\ket{Y_k}$ on $\ket{Y_{k - 1}}$, where $Y_{k + 1} = Y_k \cup Y'_k$ and $Y_{k - 1} = Y_k \cap Y'_k$ \footnote{The union of two Young diagrams is the diagram whose boxes are the union of the boxes of the two diagrams. For example, $\ydiagram{2}\cup \ydiagram{1 ,1} = \ydiagram{2, 1}$. The intersection of two Young diagrams is the diagram whose boxes are the intersection of the boxes of the two diagrams. For example, $\ydiagram{2}\cap \ydiagram{1 ,1} = \ydiagram{1}$}. For example, to compute the component of $[a_-, a_+]\ket{\ydiagram{3}}$ on $\ket{\ydiagram{2, 1}}$, it suffices to consider the component of $a_+\ket{\ydiagram{3}}$ on $\ket{\ydiagram{3, 1}}$ and the component of $a_-\ket{\ydiagram{3}}$ on $\ket{\ydiagram{2}}$. No other diagrams of size $k + 1$ and $k - 1$ contribute to this off-diagonal component of the commutator.

Now we need to apply the formula \eqref{eq: aplusminus_def} to explicitly show that off the diagonal components of $[a_-, a_+]$ are zero. To do this, we temporarily introduce the following notations for the matrix elements of $a_{\pm}$:
\begin{eqnarray}
\begin{aligned}
    a_+\ket{Y_k} = a_+(Y_{k + 1}, Y_k)\ket{Y_{k + 1}} + ...,\quad a_+\ket{Y_{k - 1}} = a_+(Y'_{k }, Y_{k - 1})\ket{Y'_{k}} + ...\\
    a_-\ket{Y_k} = a_-(Y_{k - 1}, Y_k)\ket{Y_{k - 1}} + ...,\quad a_-\ket{Y_{k + 1}} = a_-(Y'_k, Y_{k + 1})\ket{Y'_{k}} + ...
\end{aligned}
\end{eqnarray}
where $...$ denotes contributions from other basis elements which we do not need. From \eqref{eq: aplusminus_def}, the relevant coefficients are
\begin{eqnarray}\label{eq: aplus_matrix_elem}
\begin{aligned}
    a_+(Y_{k + 1}, Y_k)& = \hat{a}\left(q_1q_2 (Y_{k + 1} - Y_k) + 1 + \sum_{s\in Y_{k + 1}} \frac{Y_{k + 1} - Y_k}{s}(q_1 + q_2 -q_1q_2 - 1) \right)\ket{Y_{k + 1}}\\
    a_+(Y_{k}', Y_{k - 1}) &= \hat{a}\left(q_1q_2 (Y'_k - Y_{k - 1}) + 1 + \sum_{s\in Y'_k} \frac{Y_k' - Y_{k - 1}}{s}(q_1 + q_2 -q_1q_2 - 1) \right)\ket{Y'_{k}}\\
\end{aligned}
\end{eqnarray}
Here we identify boxes in a Young diagram with monomials in $q_1,q_2$ (appendix \ref{app: young_diag}). The notation $Y_{k + 1} - Y_k$ refers to the extra box in $Y_{k + 1}$ that is not in $Y_k$. Similar meanings are attached to other differences of Young diagrams. We also have
\begin{eqnarray}
    \begin{aligned}
    a_-(Y_{k}', Y_{k + 1}) &= \hat{a}\left(-(Y_{k + 1} - Y_k')^{-1} + 1 - \sum_{s\in Y_k'} \frac{s}{Y_{k + 1} - Y_k'}(q_1 + q_2 -q_1q_2 - 1) \right)\ket{Y_{k}'}\\
    a_-(Y_{k - 1}, Y_k) &= \hat{a}\left(-(Y_{k } - Y_{ k - 1})^{-1} + 1 - \sum_{s\in Y_{k - 1}} \frac{s}{Y_{k } - Y_{ k - 1}}(q_1 + q_2 -q_1q_2 - 1) \right)\ket{Y_{k - 1}}
    \end{aligned}
\end{eqnarray}
Therefore,
\begin{eqnarray}\label{eq: off_diagonal_comp}
    [a_-, a_+]\ket{Y_k} = \left(a_+(Y_{k + 1}, Y_k)a_-(Y_{k}', Y_{k + 1}) - a_-(Y_{k - 1}, Y_k)a_+(Y'_k, Y_{k - 1})\right)\ket{Y'_k} + ...
\end{eqnarray}
where $...$ denotes contributions from other basis vectors. We would like to show that the coefficient in front of $\ket{Y'_k}$ is zero. To do this, we need to use
\begin{eqnarray}\label{eq: extra_boxes}
\begin{aligned}
    Y_{k + 1} - Y_k = Y'_k - Y_{k - 1}\\
    Y_k - Y_{k - 1} = Y_{k + 1} - Y'_k
    \end{aligned}
\end{eqnarray}
The first line is the box in $Y'_k$ which is not in $Y_k$. The second line is the box in $Y_k$ which is not in $Y'_k$. These relations imply that
\begin{eqnarray}
\begin{aligned}
    a_+(Y_{k + 1}, Y_k) = a_+(Y'_k, Y_{k - 1})\hat{a}\left(\frac{Y_{k + 1} - Y_k}{Y_{k + 1} - Y'_k}(q_1 + q_2 - q_1q_2 - 1)\right)\\
    a_-(Y_{k - 1}, Y_k) = a_-(Y'_k, Y_{k + 1}) \hat{a}\left(\frac{Y'_k - Y_{k - 1}}{Y_k - Y_{ k - 1}}(q_1 + q_2 - q_1q_2 - 1)\right)
    \end{aligned}
\end{eqnarray}
To derive the first relation, we notice that in \eqref{eq: aplus_matrix_elem}, the only difference between the two $\hat{a}$ genera is the extra piece from $s = Y_{k + 1} - Y'_k$. The derivation of the second relation is similar. Equation \eqref{eq: extra_boxes} implies that
\begin{eqnarray}
    \frac{Y_{k + 1} - Y_k}{Y_{k + 1} - Y'_k}(q_1 + q_2 - q_1q_2 - 1) = \frac{Y'_k - Y_{k - 1}}{Y_k - Y_{ k - 1}}(q_1 + q_2 - q_1q_2 - 1)
\end{eqnarray}
Therefore, the coefficient before $\ket{Y'_k}$ in \eqref{eq: off_diagonal_comp} is zero. And $[a_-, a_+]$ is diagonal in the basis $\ket{Y}$.

\subsection{Diagonal Components of $[a_-, a_+]$}
In this section, we compute the diagonal elements of the commutator $[a_-, a_+]$. We will express $[a_-, a_+]$ as a contour integral and perform a residue calculus. First, we expand out the $\hat{a}$ genus for $a_+$ and derive a contour integral representation for $a_-$. We have
\begin{eqnarray}
\begin{aligned}
    a_+ \ket{Y} &= \sum_{z\in \text{Add}(Y)} \hat{a}(q_1q_2z)\frac{\hat{a}(q_1)\hat{a}(q_2)}{\hat{a}(q_1q_2)}\prod_{s\in Y}\frac{\hat{a}(zq_1/s)\hat{a}(zq_2/s)}{\hat{a}(zq_1q_2/s)\hat{a}(z/s)}\ket{Y + u}\\
    &= \sum_{z\in \text{Add}(Y)}\sqrt{q_1q_2}\sqrt{z}\frac{(1 -q_1^{-1}q_2^{-1}z^{-1})(1-q_1^{-1})(1-q_2^{-1})}{(1 - q_1^{-1}q_2^{-1})}\\
    &\quad\quad\prod_{s\in Y}\frac{(z/s- q_1^{-1})(z/s- q_2^{-1})}{(z/s- 1)(z/s- q_1^{-1}q_2^{-1})}\ket{Y + z}
\end{aligned}
\end{eqnarray}
The second equality follows from writing
\begin{eqnarray}
    \hat{a}(q_1^aq_2^b) = q_1^{a/2}q_2^{b/2}(1 - q_1^{-a/2}q_2^{-b/2})
\end{eqnarray}
and taking the product of the prefactors $q_1^{a/2}q_2^{b/2}$.
Similarly,\footnote{The $+1$ in the $\hat{a}$ in the definition \eqref{eq: aplusminus_def} of $a_-$ implies that the rest of the expression in $\hat{a}$ has exactly one $-1$ after simplifications. Therefore, the pole at $z = u$ is a simple pole. More work is needed to establish the $\sqrt{z}/z$ factor.} one can derive a formula for $a_-$.
\begin{eqnarray}
\begin{aligned}
    a_-\ket{Y} = \sum_{u\in \text{Rm}(Y)} \Res_{z = u} \frac{\sqrt{z}}{(z - 1)z}\prod_{s\in Y - u}\frac{(z/s - 1)(z/s - q_1q_2)}{(z/s - q_1)(z/s - q_2)}\ket{Y - u}
\end{aligned}
\end{eqnarray}
where $\Res$ denotes the residue.
As a result
\begin{eqnarray}
\begin{aligned}
    a_-a_+\ket{Y} =& \sqrt{q_1q_2}\frac{(1 - q_1^{-1})(1 - q_2^{-1})}
    {1 - q_1^{-1}q_2^{-1}}\\
    &\sum_{u\in \text{Add}(Y)}
    \Res_{z = u}\frac{1 - z^{-1}q_1^{-1}q_2^{-1}}{z - 1}\prod_{s\in Y}\Phi_{q_1, q_2}\left(\frac{z}{s}\right)\ket{Y} + ...
\end{aligned}
\end{eqnarray}
where $u$ sums over all addable boxes of $Y$ (appendix \ref{app: young_diag}) and
\begin{eqnarray}
    \Phi_{q_1, q_2}(x) = \frac{(x - q_1^{-1})(x - q_2^{-1})(x - q_1q_2)}{(x - q_1)(x - q_2)(x - q_1^{-1}q_2^{-1})}
\end{eqnarray}
is a rational function which appears in the commutation relation of the quantum toroidal algebra of $\widehat{gl_1}$ \cite{NegutElliptic}. We again used $...$ to denote contributions from other basis elements which we do not need.  A similar computation shows \footnote{The overall minus comes from shifting the product from $s\in Y- u$ to $s\in Y$ by using $\Phi_{q_1, q_2}(1) = -1$.}
\begin{eqnarray}
\begin{aligned}
    a_+a_-\ket{Y} =& -\sqrt{q_1q_2}\frac{(1 - q_1^{-1})(1 - q_2^{-1})}
    {1 - q_1^{-1}q_2^{-1}}\\
    &\sum_{u\in \text{Rm}(Y)}
    \Res_{z = u}\frac{1 - z^{-1}q_1^{-1}q_2^{-1}}{z - 1}\prod_{s\in Y}\Phi_{q_1, q_2}\left(\frac{z}{s}\right)\ket{Y}+...
\end{aligned}
\end{eqnarray}
where $u$ sums over all removable boxes of $Y$. Therefore, the commutator $[a_-, a_+]$ can be written as
\begin{eqnarray}\label{eq: commutator_contour_int}
\begin{aligned}
    [a_-, a_+]\ket{Y} =& \sqrt{q_1q_2}\frac{(1 - q_1^{-1})(1 - q_2^{-1})}
    {1 - q_1^{-1}q_2^{-1}}\\
    &\sum_{u\in \text{Add}(Y)\cup \text{Rm}(Y)}
    \Res_{z = u}z\frac{1 - z^{-1}q_1^{-1}q_2^{-1}}{(z - 1)z}\prod_{s\in Y}\Phi_{q_1,q_2}\left(\frac{z}{s}\right)\ket{Y}
\end{aligned}
\end{eqnarray}
where $u$ now sums over both the addable boxes and the removable boxes of $Y$. This contour integral can be used to prove the commutators in \cite{Schiffmann, NegutElliptic}. Hence, we can quote their results here to finish the proof. Nevertheless, we will provide a self-contained proof for the reader's convenience.

We will show that this sum of the residues is the difference between the residue at infinity and the residue at $0$. To do this, we need to show that
\begin{eqnarray}
    \frac{1 - z^{-1}q_1^{-1}q_2^{-1}}{z - 1}\prod_{s\in Y}\Phi_{q_1,q_2}\left(\frac{z}{s}\right)
\end{eqnarray}
has no pole other than $0, \infty$, the addable and the removable boxes of $Y$. To prove this, we notice that denominator vanishes at $z = sq_1, sq_2, sq_1^{-1}q_2^{-1}$. The numerator vanishes at $z = sq_1^{-1}, sq_2^{-1}, sq_1q_2$, where $s$ in a box in $Y$. Separate attention must be paid to $z = 1$ and $z = q_1^{-1}q_2^{-1}$. One can check that the only values of $u$ where there are more zeros in the denominator than these in the numerator are the ones mentioned above.
Therefore,
\begin{eqnarray}
\begin{aligned}
    \sum_{u\in \text{Add}(Y)\cup \text{Rm}(Y)}
    &\Res_{z = u}\frac{1 - z^{-1}q_1^{-1}q_2^{-1}}{z - 1}\prod_{s\in Y}\Phi_{q_1,q_2}\left(\frac{z}{s}\right)\\
    = 
    &(\Res_{z = \infty} - \Res_{z = 0})\frac{1 - z^{-1}q_1^{-1}q_2^{-1}}{z - 1}\prod_{s\in Y}\Phi_{q_1,q_2}\left(\frac{z}{s}\right)\\
    =& 1 - q_1^{-1}q_2^{-1}
\end{aligned}
\end{eqnarray}
In the last line we use the asymptotic relation $\Phi_{q_1,q_2}(0) = \Phi_{q_1,q_2}(\infty) = 1$. Now we substitute the result of this residue calculus back into \eqref{eq: commutator_contour_int} to obtain
\begin{eqnarray}
    [a_-, a_+]\ket{Y} = (\sqrt{q_1} - \sqrt{q_1}^{-1})(\sqrt{q_2} - \sqrt{q_2}^{-1})\ket{Y}
\end{eqnarray}

\section{Amplitudes of 2d $\sigma$-Models and Correspondences}
In this chapter, we show how the operators $a_\pm$ defined in the previous chapter naturally arise in 2d supersymmetric $\sigma$-models with target spaces $M_k$, the $U(1)$ instanton moduli spaces. In the next chapter, we will show a similar interpretation in terms of 6d $(1,1)$ Super Yang-Mills. We first prove that the BPS Hilbert space of the $\sigma$-model into $M_k$ has a natural basis labelled by torus fixed points of $M_k$. Therefore, the BPS Hilbert space can be identified with the domain of $a_\pm$ in the previous section. We write a BPS state as $\ket{Y_k}$ \footnote{The notation is chosen so that it is identified with $\ket{Y_k}$ in the previous chapter.} where $Y_k$ is a Young diagram which labels a torus fixed point.  States associated with different Young diagrams are orthogonal \footnote{The orthogonality can be proven using supersymmetric localisation. The complex bilinear inner product is an additional structure from physics and is not needed in the previous chapter.}. We show, using supersymmetric localisation on $[0, 1]\times S^1$ that the squared norm of these BPS states are:
\begin{eqnarray}
    \braket{Y_k}{Y_k} = \frac{1}{\hat{a}(M_k(Y_k))}
\end{eqnarray}
Next we realize the Nakajima correspondences $M_{k, k+ 1}$ in these $\sigma$-models as topological interfaces. We again work on the cylinder $[0, 1] \times S^1$ and think of the system as a string propagating in $M_k, k\geq 0$. Let $x^0$ denote the worldsheet time (the coordinate along $[0, 1]$). The topological interface is inserted at $x^0 = 0.5$ in a way that the string lies in $M_k$ when $x^0\in [0, 0.5]$ and in $M_{k + 1}$ when $x^0\in [0.5, 1]$. At $x^0 = 0.5$, the string jumps from $M_k$ to $M_{k + 1}$ via the correspondence $M_{k, k + 1}$. The topological interface naturally induces a map $a_+^{\text{2d}}$ between the BPS Hilbert space of $M_k$ and that of $M_{k + 1}$. If we prepare the BPS states $\ket{Y_k}$ at $x^0 = 0$ and $\ket{Y_{k + 1}}$ at $x^0 = 1$. The supersymmetric path integral computes the matrix element:
\begin{eqnarray}
    \bra{Y_{k + 1}}a_+^{\text{2d}}\ket{Y_k} = \frac{1}{\hat{a}(TM_{k, k + 1}(Y_k, Y_{k + 1}))}
\end{eqnarray}
These two equations implies that the topological interface realizes the operator $a_+$: 
\begin{eqnarray}
    a_+ = a_+^{\text{2d}}
\end{eqnarray}
defined by \eqref{eq: aplus_def}. A similar topological interface realizes  $a_-$.

Finally, we study how to compose correspondences from the physics point of view. We insert two topological interfaces at $x^0 = 0.3, x^0 = 0.6$ corresponding to $M_{k, k+ 1}, M_{k+1, k +2}$ respectively. Hence the string makes two successive jumps: from $M_k$ to $M_{k + 1}$ and from $M_{k + 1}$ to $M_{k + 2}$. We show that this setup correctly leads to the composition of the two correspondences $M_{k, k + 1}$ and $M_{k + 1, k + 2}$ even if they do not intersect transversely.

\subsection{Norm of BPS states}
\begin{figure}[t]
  \centering
  \includegraphics{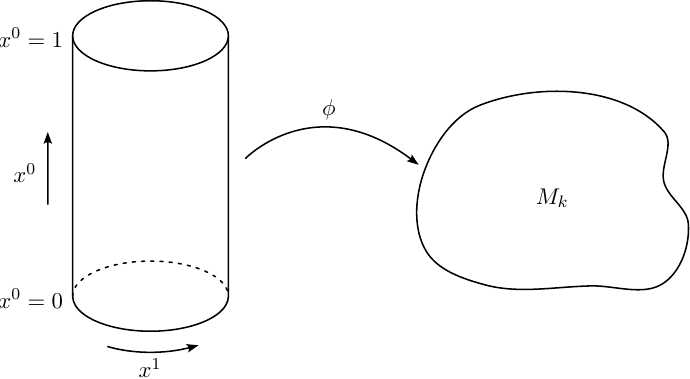}
  \caption{A closed string propagating in $M_k$.}
  \label{fig:phi}
\end{figure}

Consider a 2d $(1,1)$ supersymmetric $\sigma$-model into $M_k$, the $U(1)$ $k$-instanton moduli space. The Lagrangian is \cite{Hori2003book}
\footnote{
We suppress all inner product symbols. If $g(\cdot, \cdot)$ is the metric on $M_k$, the expression $\partial_+\phi\partial_-\phi$ means $g(\partial_+\phi, \partial_-\phi)$.
}
\begin{eqnarray}
    L[\phi, \psi_\pm, F] = 
    \partial_+ \phi \partial_- \phi 
    - \psi_+ \partial_- \psi_+ 
    - \psi_- \partial_+ \psi_- 
    + \frac{1}{2}R(\psi_+, \psi_+, \psi_-, \psi_-) 
    + FF
\end{eqnarray}
where $\phi: [0,1]\times S^1\to M_k$ (figure~\ref{fig:phi}) and
$\psi_\pm\in \phi^* TM_k$ are worldsheet fermions. $F\in \phi^*TM_{k}$ is an auxiliary boson. $R$ is the Riemann tensor of $M_k$. The subscripts $\pm$ denote the worldsheet spin. We use $x^0$ as the coordinate along $[0, 1]$ and $x^1\in [0, \beta]$ as the coordinate along $S^1$. We use the convention
\begin{eqnarray*}
    \partial_\pm = \frac{1}{2}(\partial_1 \pm i\partial_0)
\end{eqnarray*}
We also impose the twisted periodicity along $S^1$:
\begin{eqnarray}\label{eq: twisted_peri_2d}
\begin{aligned}
    &\phi(x^0, x^1 = \beta) = (q_1, q_2)\phi(x^0, x^1 = 0)\\
    &\psi_\pm(x^0, x^1 = \beta) = (q_1, q_2)_*\psi_\pm(x^0, x^1 = 0)\\
    &F(x^0, x^1 = \beta) = (q_1, q_2)_*F(x^0, x^1 = 0)
\end{aligned}    
\end{eqnarray}

Here we regard $(q_1, q_2)\in U(1)^2$ as a diffeomorphism $M_k\to M_k$ using the $U(1)^2$ action on $M_k$ and $(q_1, q_2)_*$ is the pushforward action on tangent vectors. The supersymmetry transformation laws are:
\begin{eqnarray}
\begin{aligned}
    \delta \phi = &
    \epsilon_- \psi_+ + \epsilon_+ \psi_-
    \\
    \delta F = &
    - \frac{1}{2}\epsilon_- R(\psi_+, \psi_+) \psi_-
    + \frac{1}{2}\epsilon_+ R(\psi_-, \psi_-) \psi_+  \\
    & - \epsilon_- \partial_+\psi_- 
    + \epsilon_+ \partial_- \psi_+\\
    \delta \psi_+ = &
    \epsilon_- \partial_+ \phi + \epsilon_+ F
    \\
    \delta \psi_- = &
    \epsilon_+ \partial_- \phi - \epsilon_- F
\end{aligned}
\end{eqnarray}
where $\epsilon_\pm$ are constant Grassmann parameters and $R$ is again the Riemann tensor.
The supersymmetry algebra is
\begin{eqnarray}
     \delta_{\eta_+} \delta_{\epsilon_+} = \epsilon_+ \eta_+ \partial_-
    \quad
    \delta_{\eta_-} \delta_{\epsilon_-}= \epsilon_- \eta_- \partial_+
    \quad
    [\delta_{\epsilon_+}, \delta_{\epsilon_-}] = 0
\end{eqnarray}
We impose the standard $(1, 1)$ boundary conditions preserving $\epsilon_+ = \epsilon_-$,
\begin{eqnarray}
    \partial_0\phi = F = 0
    \quad \text{and} \quad
    \psi_+ - \psi_- = 
    \partial_0\psi_+ + \partial_0 \psi_- = 0.
\end{eqnarray}
This Lagrangian is the $\delta_{\epsilon_+ = \epsilon_-}$ variation \footnote{we need to integrate by part the fermion bilinear once. Later when we insert a topological interface, it leads to a boundary terms
\begin{eqnarray}
-i\int_{x^0 = 0.5^-}\psi_-\psi_+/2 + i\int_{x^0 = 0.5^+}\psi_-\psi_+/2
\end{eqnarray}. This term must be kept in the localising action. When computing the one-loop determinant, we need to integrate by part the bosonic fluctuation which leads to the boundary term
\begin{eqnarray}
    \frac{1}{4}\int_{x^0 = 0.5^-}(\delta \phi \partial_0\delta \phi) - \frac{1}{4}\int_{x^0 = 0.5^+}(\delta \phi \partial_0\delta \phi)
\end{eqnarray}
The combination of these two terms is $\delta_{\epsilon_+ = \epsilon_-}$-exact
, where the supersymmetry now acts on the fluctuation $\delta \phi, \psi_\pm$. The supersymmetry variation of $\int_{x^0 = 0.5^-}-i\delta \phi(\psi_+ - \psi_-)/4$ gives the two boundary terms at $0.5^-$.} of
\begin{eqnarray}
    \partial_-\phi\psi_+ - F\psi_-.
\end{eqnarray}
Now we compute the overlap between BPS states.
Let $f, g$ be BPS wavefunctionals of the loop space $S^1\to M$ (together with fermions), meaning that they depend on the value of $\phi$ and $\psi_+ + \psi_-$ on a spatial circle $x^0 = \text{const}$ and are invariant under $\delta_{\epsilon_+ = \epsilon_-}$ \footnote{Note that $\phi, \psi_+ + \psi_-$ restricted to a spatial circle is invariant under the supersymmetry $\delta_{\epsilon_+ = \epsilon_-}$.}.
We would like to compute the following overlap:
\begin{eqnarray}
    \braket{f}{g} \coloneqq \int d\Phi  \exp(-S[\Phi])f(\Phi(x^0 = 1))g(\Phi(x^0 = 0))
\end{eqnarray}
where we have used $\Phi = (\phi, \psi_\pm, F)$ to denote all the fields and $\Phi(x^0 = 0, 1)$ refers to the values of the fields at the two boundary circles $x^0 = 0, 1$. $S[\Phi]$ is the integral of the Lagrangian $L[\Phi]$.
Using the Lagrangian as the localizing action, the BPS loci are constant maps to the torus fixed points. Let $Y_k$ be a fixed point. To compute the one-loop determinant, it is convenient to perform a change of variables for the fermions:
\begin{eqnarray}
    \bar{\psi} = \frac{1}{2}(\psi_+ + \psi_-)
    \quad
    \psi = \frac{1}{2}(\psi_+ - \psi_-)
\end{eqnarray}
The fermionic kinetic term becomes (up to an overall sign)
\begin{eqnarray}
    \psi_+ \partial_- \psi_+ + \psi_- \partial_+ \psi_- = 
    \bar{\psi} \partial_1 \bar{\psi}
    + \psi \partial_1 \psi
    -i \bar{\psi}\partial_0\psi
    -i \psi \partial_0 \bar{\psi},
\end{eqnarray}
which can be written in the following matrix form:
\begin{eqnarray}
    \begin{pmatrix}
        \bar{\psi} &  \psi
    \end{pmatrix}
    \begin{pmatrix}
        \partial_1 & -i\partial_0\\
        -i\partial_0 & \partial_1
    \end{pmatrix}
    \begin{pmatrix}
        \bar{\psi} \\
        \psi
    \end{pmatrix}
\end{eqnarray}
The mode matching is as follows: given a bosonic mode $\delta\phi$ which satisfies
\begin{eqnarray}
    \partial_1\delta\phi = \lambda_1\delta\phi,
    \quad
    \partial_0\delta\phi = \lambda_0\delta\phi,
\end{eqnarray}
we match it to two fermionic modes:
\begin{equation}
    \begin{pmatrix}
        \bar{\psi} \\
        \psi
    \end{pmatrix}
    \in
    \left\langle
    \begin{pmatrix}
        \delta \phi\\
        0
    \end{pmatrix},
    \begin{pmatrix}
        0\\
        \partial_0\delta\phi
    \end{pmatrix}
    \right\rangle.
\end{equation}
The matrix element of the fermionic Dirac operator is
\begin{eqnarray}
    \begin{pmatrix}
        \lambda_1 & -i\lambda_0^2 \\
        -i & \lambda_1
    \end{pmatrix}.
\end{eqnarray}
Its determinant $\lambda_1^2 + \lambda_0^2$ cancels the bosonic determinant (up to a sign). The unmatched modes of $\delta \phi, \bar{\psi}$ are constant along $x^0$ and can be expanded in Fourier modes along $x^1$. It is convenient to restrict the fluctuations to a one-complex-dimensional weight space of the fixed point with weight $q_1^{j_1} q_2^{j_2}$. We will then take the product of all the weight spaces to obtain the one-loop determinant. Within this weight space, we can treat $\delta \phi, \bar{\psi}$ as complex functions on the worldsheet and expand them in Fourier modes along $x^1$:
\begin{eqnarray}\label{eq: fourier}
    \delta \phi, \bar{\psi} \sim \exp\left(\left(\frac{2\pi in + j_1\epsilon_1 + j_2 \epsilon_2}{\beta} \right)x^1\right), n \in\Z
\end{eqnarray}
where we have temporarily set $q_1 = \exp(\epsilon_1), q_2 = \exp(\epsilon_2)$.
The one-loop determinant at the fixed point $Y_k$ is written as a product over all the weight spaces:
\begin{eqnarray}\label{eq: one_loop_det}
\begin{aligned}
    \frac{\det(\partial_1)_{\bar{\psi}}}{\det(\partial_1\partial_1)_{\delta \phi}} =& \prod_{j_1, j_2, n\in \Z}\frac{\beta}{2\pi in +  j_1 \epsilon_1 +  j_2 \epsilon_2}\\
    =& \prod_{j_1, j_2} \frac{1}{2\sinh( j_1\epsilon_1/2 + j_2\epsilon_2/2)}\\
    =& \prod_{j_1, j_2}\frac{1}{\hat{a}(q_1^{j_1}q_2^{j_2})} = \frac{1}{\hat{a}(TM_k(Y_k))}
\end{aligned}
\end{eqnarray}
where we removed a $\beta^\infty$ factor as it is a constant in our setup. We also have removed the overall square root in the left-hand side as we treat the fluctuation as complex functions.
Hence
\begin{eqnarray}
    \braket{f}{g} = \sum_{Y_k}\frac{f(Y_k)g(Y_k)}{\hat{a}(TM_k(Y_k))}
\end{eqnarray}
We define the state $\ket{Y_k}$ which evaluates to one on $Y_k$ and zero on any other tableaux $Y'_k$. Hence,
\begin{eqnarray}\label{eq: orthogonality}
    \braket{Y_k}{Y_k} = \frac{1}{\hat{a}(TM_k(Y_k))}\quad \braket{Y_k}{Y'_k} =0 \quad\forall Y_k\neq Y'_k
\end{eqnarray}
For example,
\begin{eqnarray}
    \braket{\ydiagram{1}}{\ydiagram{1}} = \frac{1}{\hat{a}(TM_1(\ydiagram{1}))} = \frac{1}{\hat{a}(q_1 + q_2)} = \frac{1}{(\sqrt{q_1} - \sqrt{q_1}^{-1})(\sqrt{q_2} - \sqrt{q_2}^{-1})}
\end{eqnarray}

\subsection{Amplitudes Associated with Correspondences}
\begin{figure}
  \centering
  \includegraphics{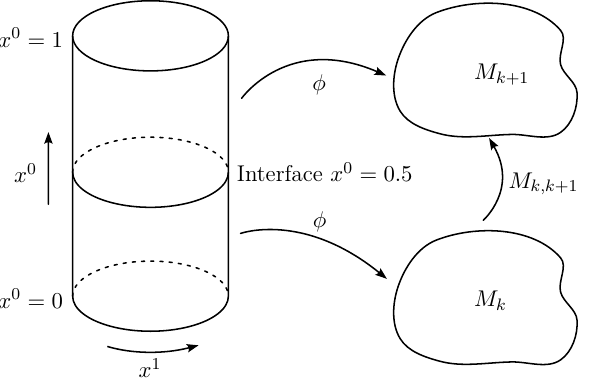}
  \caption{A topological interface inserted at $x^0 = 0.5$ (a circle) which corresponds to $M_{k, k + 1}$ (a multivalued map $M_k\to M_{k + 1}$). The number $0.5$ can be replaced by any number between $0$ and $1$.}
  \label{fig: corresp}
\end{figure}

We can generalize this computation to the case where the string jumps from $M_k$ to $M_{k + 1}$. We will consider the case where the string spends $[0,0.5]$ on $M_k$ and jumps to $M_{k + 1}$ via the correspondence at time $x^0 = 0.5$ (figure \ref{fig: corresp}).
The number $0.5$ can be replaced by any number in $(0, 1)$.
The field content is:
\begin{eqnarray}
\begin{aligned}
    &\phi: [0,0.5]\times S^1 \to M_k\\
    &\phi: [0.5,1]\times S^1\to M_{k + 1}\\
    &\psi_\pm, F\in \phi^*TM_k \quad\quad x^0\in [0, 0. 5]\\
    &\psi_\pm, F \in \phi^*TM_{k + 1}\quad x^0 \in [0.5, 1]
\end{aligned}
\end{eqnarray}
which satisfies the following conditions at the interface $x^0= 0.5$:
\begin{eqnarray}
\begin{aligned}
    &(\phi(x^0 = 0.5^-), \phi(x^0 = 0.5^+))\in M_{k, k+1}
    \\
    &(\bar{\psi}(x^0 = 0.5^-), \bar{\psi}(x^0 = 0,5^+))\in TM_{k, k + 1}
\end{aligned}
\end{eqnarray}
where $0.5^+, 0.5^-$ refer to spatial circles on the worldsheet just before $0.5$ and just after $0.5$. We have again used the notation:
\begin{eqnarray}
    \bar{\psi} = \frac{1}{2}(\psi_+ + \psi_-)
    \quad
    \psi = \frac{1}{2}(\psi_+ - \psi_-)
\end{eqnarray}
Let $f, g$ be BPS wavefunctions associated with $M_k, M_{k + 1}$ respectively. The interface at $x^0 = 0.5$ naturally leads to a map $a_+^{\text{2d}}$ from the BPS Hilbert space of $M_k$ to that of $M_{k + 1}$.  We would like to compute the following matrix element:
\begin{eqnarray}
    \bra{f}a_+^{\text{2d}}\ket{g} \coloneqq \int d\Phi \exp(-S[\Phi])f(\Phi(x^0 = 1))g(\Phi(x^0 = 0))
\end{eqnarray}
where we have used $\Phi = (\phi, \psi_\pm, F)$ to denote all the fields and $\Phi(x^0 = 0, 1)$ refers to the values of the fields at the two boundary circles $x^0 = 0, 1$.
The localization argument in the previous section shows that the path integral localizes to piecewise constant maps
\begin{eqnarray}\label{eq: bps_locus_2d}
    \phi(x^0\in [0, 0. 5]) \equiv Y_k\quad \phi(x^0 \in [0. 5, 1]) \equiv Y_{k + 1} \quad (Y_k, Y_{k + 1}) \in M_{k, k +1}
\end{eqnarray}
where $Y_k, Y_{k + 1}$ are torus fixed points such that $Y_k\subset Y_{k + 1}$.
The mode matching in the previous section is still valid due to the boundary conditions \footnote{One needs to be careful with the boundary terms at the interface coming from integration by part. See the first footnote in the previous section for more information.}. The unmatched modes $\delta \phi, \delta \bar{\psi}$ are piecewise constant along $x^0$ (constant in the interval $[0, 0. 5]$ and $[0. 5, 1]$ separately) and can be expanded in Fourier modes along $x^1$ in a way similar to \eqref{eq: fourier}. The interface condition at $x^0 = 0.5$ forces $(\delta \phi(x^0 = 0.5^-), \delta \phi(x^0 = 0.5^+))$ and $(\delta \bar{\psi}(x^0 = 0.5^-), \delta \bar{\psi}(x^0 = 0.5^+))$ to be tangent to $M_{k, k + 1}$. The resulting one-loop determinant at $(Y_k, Y_{k + 1})$ is
\begin{eqnarray}
    \frac{1}{\hat{a}(TM_{k, k + 1}(Y_k, Y_{k + 1}))}
\end{eqnarray}
The final result for the amplitude is
\begin{eqnarray}
    \bra{f}a_+^{\text{2d}}\ket{g} = \sum_{Y_k\subset Y_{k + 1}}
    \frac{f(Y_{k + 1})g(Y_k)}{\hat{a}(TM_{k, k + 1}(Y_k, Y_{k + 1}))}
\end{eqnarray}
In particular,
\begin{eqnarray}
    \bra{Y_{k + 1}}a_+^{\text{2d}}\ket{Y_k} =
    \frac{1}{\hat{a}(TM_{k, k + 1}(Y_k, Y_{k + 1}))}
\end{eqnarray}
This equation and \eqref{eq: orthogonality} imply that $a_+$ defined in \eqref{eq: aplus_def} and $a_+^{\text{2d}}$ have the same matrix coefficients. Hence, they are equal: 
\begin{eqnarray}
    a_+ = a_+^{\text{2d}}
\end{eqnarray}
and we have successfully realized $a_+$ as topological interfaces in the 2d $\sigma$-models.
A similar argument allows us to also realize $a_-$ as topological interfaces in these 2d $\sigma$-models.
\subsection{Composition of Correspondences}
In this section, we realize the compositions of correspondences in physics by inserting two topological interfaces at the two circles $x^0 = 0.3$ and $x^0 = 0.6$. In other words, the string jumps from $M_k$ to $M_{k + 1}$ at $x^0 = 0.3$ via $M_{k, k+ 1}$ and makes a second jump from $M_{k + 1}$ to $M_{k + 2}$ at $x^0 = 0.6$ via $M_{k + 1, k+ 2}$. We will show that supersymmetric localisation correctly produces the composition of the correspondences even when they do not intersect transversely. This section is not needed for the rest of this work, but we include it because it illustrates the power of supersymmetric localisation \footnote{The computations we perform here is similar to the computations of the Witten index of an open string stretched between two B-branes. The excess bundle we will encounter also arises when the two B-branes do not intersect transversely.}.

\begin{figure}
  \centering
  \includegraphics{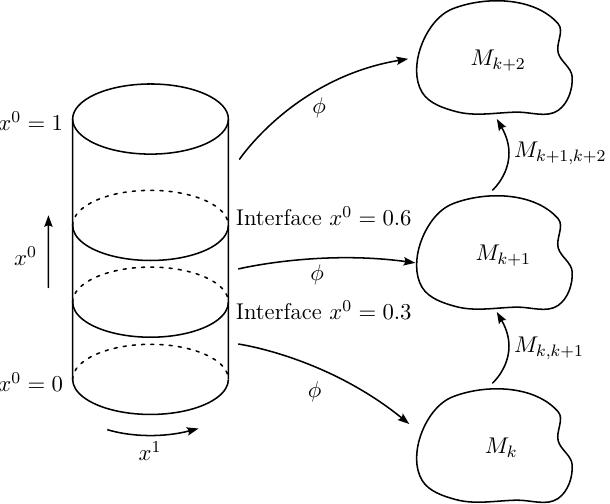}
  \caption{Two topological interfaces inserted at $x^0 = 0.3, 0.6$ which correspond to $M_{k, k + 1}, M_{k + 1, k +2 }$ respectively. The numbers $0.3, 0.6$ can be replaced by any two distinct numbers between $0$ and $1$.}
  \label{fig: two_corresp}
\end{figure}

We insert two topological interfaces at  corresponding to $M_{k, k+ 1}$ and $M_{k+ 1, k +2 }$ respectively (figure \ref{fig: two_corresp}).
The number $0.3, 0.6$ can be replaced by any two distinct numbers between 0 and 1.
The field content is
\begin{equation}
\begin{aligned}
    &\phi: \left[0, 0.3\right]\times [0,\beta] \to M_k\\
    &\phi: \left[0.3, 0.6\right]\times [0, \beta]\to M_{k + 1}\\
    &\phi: \left[0.6, 1\right]\times [0, \beta]\to M_{k + 2}\\
    &\psi_\pm, F\in \phi^*TM_k \quad\quad x^0\in \left[0, 0.3\right]\\
    &\psi_\pm, F \in \phi^*TM_{k + 1}\quad x^0 \in \left[0.3, 0.6\right]\\
    &\psi_\pm, F \in \phi^*TM_{k + 2}\quad x^0 \in \left[0.6, 1\right]
\end{aligned}
\end{equation}

We impose the same boundary conditions at $x^0 = 0, 1$ as before and the same interface conditions at $x^0 = 0.3, 0.6$.  Let $f, g$ be BPS wavefunctionals associated with $M_k, M_{k +2}$ respectively.
We would like to compute the amplitude \footnote{We regard $a_+^{\text{2d}}\circ a_+^{\text{2d}}$ as an operator defined by this path integral. Later we will prove that $a_+^{\text{2d}}\circ a_+^{\text{2d}}$ is the composition of $a_+^{\text{2d}}$ with itself.}
\begin{eqnarray}\label{eq: amplit_two_corresp}
    \bra{f}a_+^{\text{2d}}\circ a_+^{\text{2d}}\ket{g} \coloneqq \int d\Phi \exp(-S[\Phi])f(\Phi(x^0 = 1))g(\Phi(x^0 = 0))
\end{eqnarray}
where we have again used $\Phi = (\phi, \psi_\pm, F)$ to denote the entire motion of the particle.

The path integral localizes to piecewise constant maps
\begin{eqnarray*}
    \phi(x^0 \in [0, 0.3]) \equiv Y_1\quad \phi(x^0 \in [0.3, 0.6]) \equiv Y_2 \quad \phi(x^0 \in [0.6, 1]) \equiv Y_3
\end{eqnarray*}
where
\begin{eqnarray}
    (Y_1, Y_2) \in M_{k, k + 1} \quad (Y_2, Y_3)\in M_{k + 1, k + 2}
\end{eqnarray}
 are torus fixed points. Equivalently,
 \begin{eqnarray}
     (Y_1, Y_2, Y_3)\in M_{k, k + 1, k + 2}
 \end{eqnarray}
where
\begin{eqnarray}
    M_{k, k + 1, k + 2} = (M_{k, k + 1}\times M_{k + 2})\cap (M_k\times M_{k + 1, k + 2})\subset M_k\times M_{k + 1}\times M_{k + 2}
\end{eqnarray}
is the set-theoretic intersection \footnote{The set-theoretic intersection may be singular. When that is the case, we define the tangent space of $M_{k, k+1,k+2}$ at a fixed point to be:
\begin{eqnarray}
    TM_{k, k + 1, k +2} = (TM_{k, k + 1} \oplus TM_{k + 2}) \cap (TM_{k}\oplus TM_{k+1, k+ 2})
\end{eqnarray}
We only need the tangent space $TM_{k, k + 1, k +2}$  at a fixed point (not the global structure of $M_{k, k +1, k +2}$) to perform the localisation computation at that point.
} of the two correspondences which does not need to be transverse. 

The mode matching in the previous section is still valid. We now consider the unmatched modes:
\begin{enumerate}
    \item[(a)] The unmatched modes of $\delta \phi, \bar{\psi}$ consist of piecewise constant (along $x^0$) modes so that
    \begin{eqnarray}
    \begin{aligned}
        &(\delta \phi(x^0 = 0.3^-), \delta \phi(x^0 = 0.3^+)\in TM_{k, k +1}\\
        &(\delta \phi(x^0 = 0.6^-), \delta \phi(x^0 = 0.6^+)\in TM_{k + 1, k +2}
    \end{aligned}
    \end{eqnarray}
     Therefore, unmatched modes of $\delta \phi$ biject with tangent vectors of $M_{k, k +1, k+2}$ times Fourier modes along $x^1$ as in \eqref{eq: fourier}. The unmatched modes of $\bar{\psi}$ are the same as those of $\delta \phi$ and they contribute
    \begin{eqnarray}
        \frac{1}{\hat{a}(TM_{k, k+1, k+2}(Y_k, Y_{k + 1}, Y_{k + 2}))}
    \end{eqnarray}
    to the one-loop determinant.
    \item[(b)] The unmatched modes of $\psi$ are orthogonal to $\partial_0\delta \phi$ for all $\delta \phi$. Therefore, $\psi$ must be piecewise constant along $x^0$. The boundary conditions $\psi(x^0 = 0) = \psi(x^0 = 1) = 0$ force $\psi$ to be 0 on $[0,0.3]\cup [0.6,1]$. On $[0.3, 0.6]$, we claim that $\psi$ must be orthogonal to $TM_{k, k + 1}$ and $TM_{k+1, k +2}$.
    \footnote{
        More precisely, $\psi$ is orthogonal to the images of the natural projection maps $TM_{k, k +1}(Y_k, Y_{k + 1})\to TM_{k + 1}(Y_{k + 1})$ and $TM_{k + 1, k +2}(Y_{k + 1}, Y_{k + 2})\to TM_{k + 1}(Y_{k + 1})$.
    } To prove this statement, consider $\delta \phi$ which are supported in a small neighborhood of $x^0 = 0.3^+$. The equality
    \footnote{
        We remind the reader of our convention of not writing out the inner product symbol.
    }
    \begin{eqnarray}
        \int_{[0, 1]\times S^1} \partial_0\delta \phi \psi= - \int_{x^0 = 0.3^+}\delta \phi(0.3^+)\psi(0.3^+)
    \end{eqnarray}
    forces $\psi(0.3^+)$ to be orthogonal to $TM_{k, k + 1}$. A similar argument using $\delta \phi$ supported around $x^0 = 0.6^-$ shows that $\psi(0.6^-)$ is orthogonal to $M_{k + 1, k +2}$. The unmatched modes of $\psi$ are constant along $x^0$ in the interval $[0.3, 0.6]$. Hence $\psi(0.3^+) = \psi(0.6^-)$ is orthogonal to both $TM_{k, k + 1}$ and $TM_{k + 1, k + 2}$ and takes value in the excess bundle
    \begin{eqnarray}
        EM_{k, k + 1, k +2} = \frac{TM_k\oplus  TM_{k + 1}\oplus TM_{k + 2}}{(TM_{k, k + 1}\oplus TM_{k + 2}) + (TM_k\oplus TM_{k + 1, k +2})}
    \end{eqnarray}
    of the intersection $(M_k \times M_{k + 1, k + 2})\cap (M_{k , k+ 1}\times M_{k + 2})$. It measures the extent away from a transverse intersection. If the intersection is transverse, $EM_{k , k+ 1, k+2} = 0$.
    
    Therefore, the unmatched modes of $\psi$ contributes
    \begin{eqnarray}
        \hat{a}(EM_{k, k+ 1, k + 2}(Y_k, Y_{k + 1}, Y_{k + 2}))
    \end{eqnarray}
    to the one-loop determinant.
\end{enumerate}
The full amplitude is
\begin{eqnarray}
\begin{aligned}
    &\bra{f}a_+^{\text{2d}}\circ a_+^{\text{2d}}\ket{g} \\
    =& \sum_{Y_k\subset Y_{k + 1}\subset Y_{k + 2}}f(Y_{k + 2})g(Y_k)\frac{\hat{a}(EM_{k, k + 1, k +2}(Y_k, Y_{k + 1}, Y_{k + 2}))}{\hat{a}(TM_{k, k + 1, k +2}(Y_k, Y_{k + 1}, Y_{k + 2}))
    }
\end{aligned}
\end{eqnarray}
In particular,
\begin{eqnarray}
    \bra{Y_{k + 2}}a_+^{\text{2d}}\circ a_+^{\text{2d}}\ket{Y_k} = \sum_{Y_k\subset Y_{k + 1}\subset Y_{k + 2}}\frac{\hat{a}(EM_{k, k + 1, k +2}(Y_k, Y_{k + 1}, Y_{k + 2})}{\hat{a}(TM_{k, k + 1, k +2}(Y_k, Y_{k + 1}, Y_{k + 2})
    }
\end{eqnarray}
where we sum over all $Y_{k + 1}$ which contains $Y_k$ and is contained in $Y_{k  + 2}$.

Now we prove that operator $a_+^{\text{2d}}\circ a_+^{\text{2d}}$ defined using the supersymmetric path integral \eqref{eq: amplit_two_corresp} is the composition of the two operators $a_+^{\text{2d}}a_+^{\text{2d}}$. We will suppress the dependence on the fixed point and write $TM_{k , k + 1}(Y_k, Y_{k + 1}) = TM_{k , k + 1}$ to improve the readability. The derivation goes as follows:
\begin{eqnarray}
\begin{aligned}
    &\bra{Y_{k + 2}}a_+^{\text{2d}}a_+^{\text{2d}}\ket{Y_k}\\
    =& \sum_{Y_{k + 1}}\frac{\bra{Y_{k + 2}}a_+^{\text{2d}}\ket{Y_{k + 1}}\bra{Y_{k + 1}}a_+^{\text{2d}}\ket{Y_k}}{\braket{Y_{k + 1}}{Y_{k + 1}}}\\
    =& \sum_{Y_{k + 1}} \frac{\hat{a}(TM_{k + 1})}{\hat{a} (TM_{k, k + 1}) \hat{a} (TM_{k + 1, k + 2})}\\
    = &\sum_{Y_{k + 1}} \frac{
        \hat{a}(TM_{k + 1}\oplus TM_k \oplus TM_{k + 2})
        }{
            \hat{a} (TM_{k, k + 1} \oplus TM_{k + 2}) \hat{a} (TM_{k + 1, k + 2} \oplus TM_k)
        }\\
    =& \sum_{Y_{k + 1}} \frac{\hat{a}(EM_{k, k+ 1, k + 2}(Y_k, Y_{k + 1}, Y_{k + 2}))}{\hat{a}(TM_{k, k+ 1, k + 2}(Y_k, Y_{k + 1}, Y_{k + 2}))}\\
    =& \bra{Y_{k + 2}}a_+^{\text{2d}}\circ a_+^{\text{2d}}\ket{Y_k}
\end{aligned}
\end{eqnarray}
where $Y_{k + 1}$ sums over all Young diagrams with $k + 1$ boxes which contains $Y_k$ and is contained in $Y_{k  + 2}$.
In the last step we used the fact that if $V_1, V_2\subset V$ are subrepresentations of a representation $V$ of $U(1)^2$, then the characters satisfy
\begin{eqnarray}
    \ch(V_1) + \ch(V_2) -\ch(V) = \ch(V_1 \cap V_2) - \ch(V/(V_1\cup V_2))
\end{eqnarray}
and we set
\begin{eqnarray}
\begin{aligned}
    V_1 =& TM_{k, k +1}\oplus TM_{k + 2}\\ 
    V_2 =& TM_{k+1, k + 2}\oplus TM_k\\
    V =& TM_k \oplus TM_{k + 1} \oplus TM_{k + 2} \\
    V_1 \cap V_2 =& TM_{k, k +1,k +2}
    \\ V/(V_1 \cap V_2) =& EM_{k, k +1, k +2}
\end{aligned}
\end{eqnarray}
As a result, the two topological interfaces correctly realize the composition of the operators $a_+a_+$.

\section{Instanton Line Operators in 6d $(1, 1)$ Super Yang-Mills}
In this chapter, we realize the 2d-$\sigma$ models into instanton moduli spaces in the previous chapter as low energy dynamics of 6d $(1, 1)$ Super Yang-Mills on $\C^2 \times [0, 1]\times S^1$ with a twisted periodicity along $S^1$. The topological interface associated with the Nakajima correspondences $M_{k, k+ 1}$ corresponds to instanton line operators $a_\pm^{\text{6d}}$ of charge $\pm 1$ wrapping the $S^1$ factor. These operators are defined by integrating over gauge fields with Chern class $\pm 1$ on a four sphere surrounding the line in the path integral.  We will perform supersymmetric localisation directly in the 6d gauge theory to reproduce the formula \eqref{eq: matrix_elem} \eqref{eq: overlap}. First we explain how to perform a topological twist to preserve two supercharges. We then write down the off-shell Lagrangians and half-BPS boundary conditions. Finally, we provide a detailed computation of the one-loop determinant.
\subsection{Topological Twist}
\begin{figure}
  \centering
  \includegraphics{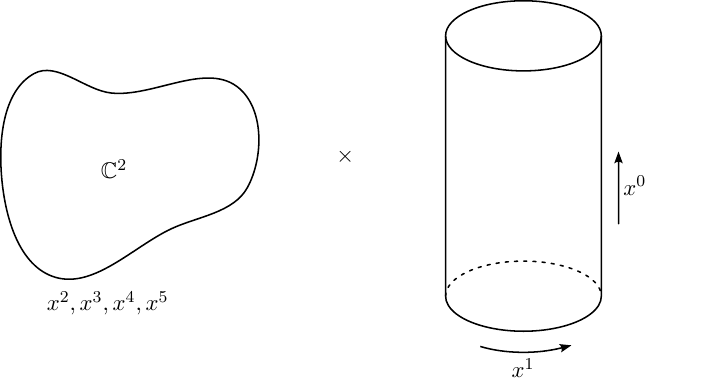}
  \caption{6d $(1, 1)$ super Yang-Mills on $\C^2$ times a cylinder}
  \label{fig:m_times_cylinder}
\end{figure}

We place 6d $U(1)$ $(1,1)$ Super Yang-Mills on the flat spacetime $\C^2 \times [0, 1] \times S^1$. We use $x^0$ for the time coordinate along $[0, 1]$ and $x^1\in [0, \beta]$ as the coordinate along $S^1$. We also use real flat coordinates $x^\mu, \mu = 2, 3,4,5$ along $\C^2$. They are related to the natural complex coordinates $z_1, z_2$ via
\begin{eqnarray}
    z_1 = x^2 + ix^3 \quad z_2 = x^4 + ix^5
\end{eqnarray}(figure~\ref{fig:m_times_cylinder}).
The natural $U(1)^2$ action on $\C^2 \times [0, 1]$ given by
\begin{eqnarray}
    (z_1, z_2, x^0)\mapsto (q_1z_1, q_2z_2, x^0)\quad (q_1, q_2)\in U(1)^2
\end{eqnarray}
induces an action on classical fields living on $\C^2 \times [0, 1]$. This action allows us to switch on the twisted periodicity along $S^1$:
\begin{eqnarray}\label{eq: twisted_peri_6d}
    \Phi(x^0, x^\mu, x^1 = \beta) = (q_1, q_2)_*
    \Phi(x^0, x^\mu, x^1 = 0)
\end{eqnarray}
In other words, $\Phi$ restricted to the five dimensional slice $x^0 = \beta$ is a rotated version of $\Phi$ restricted to the slice $x^0 = 0$, rotated by $(q_1, q_2)\in U(1)^2$.

This twisted periodicity breaks supersymmetry.
To preserve supersymmetry, we will perform a topological twist along $\C^2$ using a map $SO(4)_{\C^2}\to SO(4)_R$, where $SO(4)_{\C^2}$ is the local rotation group of $\C^2$ and $SO(4)_R$ is the R-symmetry of the 6d $(1, 1)$ super Yang-Mills. Under the decomposition,
\begin{equation}
    SO(4)_{\C^2} \cong SU(2)_{\C^2}^+ \times SU(2)_{\C^2}^-
    \quad 
    SO(4)_R \cong SU(2)_R^+ \times SU(2)_R^-
\end{equation}
the topological twist uses the diagonal embedding $SU(2)_{\C^2}^+\to SU(2)_R^+\times SU(2)_R^-$ \footnote{There is another twist which uses the identity map $SO(4)_{\C^2}\to SO(4)_R$ and produces two anticommuting scalar supercharges with the same weight under $SO(2)_\pm$. The low energy dynamics on the cylinder is a 2d (0, 2) $\sigma$-model into the tangent bundles of instanton moduli spaces parametrized by $A_\mu$ and the bosonic vector $\phi_\mu$. This space seems to admit a complex structure. A third twist which uses the map $SU(2)^+_{\C^2}\to SU(2)^+_R$ is equivalent to the Donaldson-Witten of an $\mathcal{N} = 2$ $U(N)$ vector multiplet with an adjoint hypermultiplet. The hypermultiplet scalars become spinors on $\C^2$ and the low-energy dynamics is a 2d (0, 1) $\sigma$-model into the instanton moduli spaces on $\C^2$ coupled to the bundles of Dirac zero modes. We will not study these two twists in this paper.}. Half of the 16 supercharges of the 6d theory transform under $SU(2)_R^+$ and the other half transforms under $SU(2)_R^-$. Therefore, we obtain two independent Donaldson-Witten twists \cite{WittenMonopole}, one for each half of the supercharges. We obtain two scalar supercharges $Q_\pm$ (with respect to $\C^2$) with opposite spins under the $SO(2)_{\pm}$ local rotation group of the cylinder. Their commutators are \footnote{We have not written out gauge transformation factors here. They will be included in the formulas later.} 
\begin{equation} \label{6d_susy_algebra}
    Q_+^2 = \partial_+ \quad Q_-^2 = \partial_- \quad \{Q_+, Q_-\} = 0
\end{equation}
We use the following convention which is the same as the convention for the 2d $\sigma$-model in the previous chapter:
\begin{eqnarray}
    \partial_\pm = \frac{1}{2}(\partial_1 \pm i\partial_0)
\end{eqnarray}

\subsection{Supersymmetry Transformation Laws}
The topological twist of the 6d $(1,1)$ gauge theory in the previous section changes the spins of the fields. We write the field content after the twist as: 
\begin{eqnarray}
    \begin{aligned}
        &A_\mu \text{ (gauge field along } \C^2)\\
    &A_\pm \text{ (gauge field along the cylinder})\\
    &\phi \text{  (bosonic scalar)}\\
    &\phi^+_{\mu\nu} \text{ (bosonic self-dual two-form)}\\
    &\lambda, \bar{\lambda} \text{ (fermionic scalars)}\\
    &\lambda_\mu, \bar{\lambda}_\mu \text{ (fermionic one-forms)}\\
    &\lambda^+_{\mu\nu}, \bar{\lambda}^+_{\mu\nu} \text{ (fermionic self-dual two-forms)}
    \end{aligned}
\end{eqnarray}
    
The terms ``one-form'' and ``self-dual two-form'' refer to the transformation laws under $SO(4)_{\C^2}$, the local rotation group of $\C^2$. The overlines on the fermions denote positive weights under $SO(2)_\pm$, the local rotation group of the cylinder (parametrized by $x^0, x^1$). We also add two bosonic auxiliary fields \footnote{It would be interesting to know whether the seven auxiliary fields are the same auxiliary fields in \cite{Berkovits1993}.}:
\begin{eqnarray}
    \begin{aligned}
        &H_\mu \text{ (one-form auxiliary field)}\\
    &H^+_{\mu\nu} \text{ (self-dual two-form auxiliary field)}
    \end{aligned}
\end{eqnarray}
    
The abelian \footnote{The formulas for the nonabelian cases are longer, especially for the auxiliary fields.} supersymmetry transformation laws are \footnote{Note the symmetry that exchanges fermions with and without overlines, $\epsilon$ and $\bar{\epsilon}$, $\partial_+$ and $\partial_-$, $A_+$ and $A_-$, $H_\mu$ and $-H_\mu$, $H^+_{\mu\nu}$ and $-H^+_{\mu\nu}$. This is the statement that 6d (1,1) is non-chiral.}:
\begin{equation}
    \begin{aligned}
        \delta A_+& = i\epsilon \lambda \quad \delta A_- = i\bar{\epsilon}\bar{\lambda}\\
    \delta
    \begin{pmatrix}
        A_\mu\\
        \phi\\
        \phi^+_{\mu\nu}
    \end{pmatrix} &= \epsilon 
    \begin{pmatrix}
        \bar{\lambda}_\mu\\
        \bar{\lambda}\\
        \bar{\lambda}^+_{\mu\nu} 
    \end{pmatrix}+ \bar{\epsilon}
    \begin{pmatrix}
        \lambda_\mu\\
        \lambda\\
        \lambda^+_{\mu\nu}
    \end{pmatrix}\\
    \delta \begin{pmatrix}
        \lambda_\mu\\
        \lambda\\
        \lambda^+_{\mu\nu}
    \end{pmatrix} &= i\bar{\epsilon} (\partial_+ + A_+)
    \begin{pmatrix}
        A_\mu\\
        \phi\\
        \phi^+_{\mu\nu}
    \end{pmatrix} + \epsilon
    \begin{pmatrix}
        -\frac{1}{2}\partial_\mu\phi - \frac{1}{2}H_\mu \\
        -F_{+-}\\
         -  \frac{1}{2}H_{\mu\nu}^+
    \end{pmatrix}\\
    \delta \begin{pmatrix}
        \bar{\lambda}_\mu\\
        \bar{\lambda}\\
        \bar{\lambda}^+_{\mu\nu}
    \end{pmatrix} &= i\epsilon (\partial_- + A_-)
    \begin{pmatrix}
        A_\mu\\
        \phi\\
        \phi^+_{\mu\nu}
    \end{pmatrix} + \bar{ \epsilon}
    \begin{pmatrix}
        -\frac{1}{2}\partial_\mu\phi + \frac{1}{2}H_\mu \\
        -F_{-+}\\
         + \frac{1}{2}H_{\mu\nu}^+
    \end{pmatrix}\\
    \delta H_\mu &= -2 i \epsilon \partial_- \lambda_\mu - \epsilon \partial_\mu\bar{\lambda} +2i \bar{\epsilon}\partial_+\bar{\lambda}_\mu + \bar{\epsilon}\partial_\mu \lambda \\
    \delta H_{\mu\nu}^+ &= -2i \epsilon \partial_- \lambda_{\mu\nu}^+ + 2 i \bar{\epsilon}\partial_+ \bar{\lambda}_{\mu\nu}^+
    \end{aligned}
\end{equation}
The supersymmetry algebra is:
\begin{eqnarray}
    \begin{aligned}
        \delta_\eta\delta_{\epsilon} = i\epsilon\eta(\partial_- + A_-) \quad \delta_{\bar{\eta}}\delta_{\bar{\epsilon}} = i\bar{\epsilon}\bar{\eta}(\partial_+ + A_+) \quad [\delta_{\bar{\epsilon}}, \delta_\epsilon] = \epsilon\bar{\epsilon}\phi
    \end{aligned}
\end{eqnarray}
The actions of gauge transformations are
\begin{eqnarray*}
    \delta A_\mu = -\partial_\mu C \quad \delta A_\pm = -\partial_\pm C,
\end{eqnarray*}
where $C$ is a gauge transformation parameter. For example,
\begin{eqnarray*}
    (\partial_+ + A_+) A_\mu = \partial_+ A_\mu - \partial_\mu A_+ = F_{+\mu}.
\end{eqnarray*}
\subsection{Lagrangians and Boundary Conditions}
The full Lagrangian is both $\delta_\epsilon$ and $\delta_{\bar{\epsilon}}$ exact (modulo total derivative terms) \footnote{The sign in front of $\phi^+_{\mu\nu}F_{\mu\nu}$ is convention-dependent. It does not change the final Lagrangian after integrating out the auxiliary fields.}:
\begin{eqnarray}
    \begin{aligned}
    -\epsilon\bar{\epsilon}L = \delta_{\epsilon}\delta_{\bar{\epsilon}}\left[ (\bar{\lambda}_\mu, \bar{\lambda}, \bar{\lambda}^+_{\mu\nu})
    \begin{pmatrix}
        \lambda_\mu\\
        \lambda\\
        \lambda^+_{\mu\nu}
    \end{pmatrix}
    + \phi^+_{\mu\nu} F_{\mu\nu} \right]
    \end{aligned}
\end{eqnarray}
The bosonic part of the Lagrangian is
\begin{eqnarray}
    \begin{aligned}
        &F_{+\mu} F_{-\mu} + \partial_+\phi \partial_-\phi + \partial_+ \phi^+_{\mu\nu} \partial_-\phi^+_{\mu\nu} - \frac{1}{4} H_{\mu\nu}^+ H_{\mu\nu}^+ - \frac{1}{4}H_\mu H^\mu \\
    &+ F_{+-}F_{-+}
    + \frac{1}{4}D_\rho \phi D^\rho \phi - \frac{1}{2}H^+_{\mu\nu}F^+_{\mu\nu} + \partial^\mu \phi^+_{\mu\nu}H^\nu
    \end{aligned}
\end{eqnarray}

and the fermionic part is
\begin{eqnarray}
    \begin{aligned}
        &i(\bar{\lambda}_\mu, \bar{\lambda}, \bar{\lambda}^+_{\mu\nu})\partial_+
    \begin{pmatrix}
        \bar{\lambda}_\mu\\
        \bar{\lambda}\\
        \bar{\lambda}^+_{\mu\nu}
    \end{pmatrix}
    +
    i(\lambda_\mu, \lambda, \lambda^+_{\mu\nu})\partial_-
    \begin{pmatrix}
        \lambda_\mu\\
        \lambda\\
        \lambda^+_{\mu\nu}
    \end{pmatrix}\\
    &+ \lambda_\mu \partial^\mu \bar{\lambda} + \bar{\lambda}_\mu \partial_\mu \lambda +2 \bar{\lambda}^+_{\mu\nu}\partial_\mu \lambda_\nu - 2\lambda^+_{\mu\nu}\partial_\mu \bar{\lambda}_\nu.
    \end{aligned}
\end{eqnarray}
We impose the following half BPS boundary conditions preserving $\epsilon = \bar{\epsilon}$ \footnote{Analogous boundary conditions for 3d gauge theories were studied extensively in \cite{DimofteDualB, BullimoreBMS}. Boundary conditions for 4d maximal super Yang-Mills were studied in \cite{GaiottoSdualityN=4}. It would be interesting to know if the dimensional reduction of our boundary conditions match the boundary conditions in \cite{GaiottoSdualityN=4}.}:
\begin{eqnarray}\label{6d_neumann}
\begin{aligned}
    &A_0 = \partial_0A_1 = \partial_0 A_\mu  = \partial_0 \phi  = \phi^+_{\mu\nu} = \partial_0\partial_0\phi^+_{\mu\nu}- \partial_0H^+_{\mu\nu} = H_\mu =  0\\
    &\bar{\lambda}_\mu - \lambda_\mu = \lambda - \bar{\lambda} = \partial_0(\bar{\lambda}^+_{\mu\nu} - \lambda^+_{\mu\nu}) = 0 \\
    &\partial_0(\bar{\lambda}_\mu + \lambda_\mu) = \partial_0(\lambda + \bar{\lambda}) = \bar{\lambda}^+_{\mu\nu} + \lambda^+_{\mu\nu} = 0
\end{aligned}
\end{eqnarray}
The Lagrangian is the $\delta_{\epsilon = \bar{\epsilon}}$ variation of the following:
\begin{eqnarray}
    \begin{aligned}
        &F_{\mu\nu}\lambda^+_{\mu\nu} + 2\phi^+_{\mu\nu} \partial_\mu \lambda_\nu - iF_{+\mu}\bar{\lambda}_\mu -\frac{1}{2} \lambda_\mu (\partial_\mu \phi - H_\mu)\\
    &+ \frac{1}{2}H^+_{\mu\nu}\lambda^+_{\mu\nu} - i\bar{\lambda}^+_{\mu\nu}\partial_+\phi^+_{\mu\nu} - F_{-+}\lambda - i\bar{\lambda}\partial_+\phi
    \end{aligned}
\end{eqnarray}
We will use the Lagrangian as the localising action.

\subsection{Localization}
In this section, we compute the matrix elements of the charge one instanton line operator $a_+^{\text{6d}}$ using localisation. Let $f, g$ be BPS wavefunctionals of the 6d gauge theories. We would like to compute the following overlap \footnote{Analogous localisation computations were performed for monopole operators in \cite{Tamagni2024}.}
\begin{eqnarray}
    \bra{f}a_+^{\text{6d}}\ket{g} \coloneqq \int f(\Phi(x^0 = 1)) g(\Phi(x^0 = 0)) \exp(- S[\Phi])
\end{eqnarray}
where $\Phi$ denote all the fields in the 6d gauge theory and $\Phi(x^0 = 0, 1)$ denote the values of these fields on the two boundaries $x^0 = 0, x^0 = 1$. We insert the charge one instanton line operator at the origin of $\C^2$,  $x^0 = 0.5$ and fills out the $S^1$ factor. Therefore, the instanton number along $\C^2$ at $x^0 = 0.5^+$ is one bigger than the instanton number at $x^0 = 0.5^-$.
The minimum of the action is attained at
\begin{eqnarray}
    \phi = \phi^+_{\mu\nu} = F_{\mu\nu}^+ = F_{\mu \pm} = F_{+-} = 0.
\end{eqnarray}
These equations imply that $A_\mu$ are $U(1)$ instantons along each time slice. We can also set $A_\pm = 0$ by a gauge transformation \footnote{The proof of this statement requires the extra condition that $A$ has vanishing holonomy along the $S^1$. Therefore, we impose the condition that the holonomy of $A$ along $S^1$ tends to zero at the infinity of $\C^2$ from the beginning. At the BPS locus, $F_{\mu \pm}$ forces the holonomy to vanish everywhere. A flat connection on $S^1 \times [0,1]$ with zero $S^1$ holonomy is always a pure gauge. $\C^2$ does not impose any topological obstruction to the global choice of the gauge transformation parameter.}. The twisted periodicity \eqref{eq: twisted_peri_6d} implies that these instantons are torus fixed points. Hence $A_\mu = Y_k$ when $x^0\leq 0.5$ and $A_\mu = Y_{k + 1}$ when $x^0\geq 0.5$. As the instanton operator is located at the origin, we further impose $Y_k\subset Y_{k + 1}$ \footnote{A better argument for this claim would be to analyze the low energy dynamics without the twisted periodicity along $S^1$. One can show the BPS locus in this case is exactly $M_{k , k+ 1}$.}. The 6d BPS locus is the same as the 2d BPS locus \eqref{eq: bps_locus_2d}. 

Next we compute the one-loop determinant at a fixed point $(Y_k, Y_{k + 1})$\footnote{The intermediate steps in this localisation are not rigorous as we are working with $U(1)$ instantons. We expect that our computations can be made more rigorous by switching on a noncommutative deformation of $\C^2$ \cite{NekrasovNoncommutative}.}. First we gauge fix by adding the gauge fixing terms \footnote{We gauge fix the one-loop fluctuation, not the original Lagrangian.}
\begin{eqnarray}
    \frac{1}{4}\left(\sum_{i = 0}^5 \partial_i \delta A^i\right)^2 - \sum_{i=0}^5\bar{C}\partial_i\partial^i C.
\end{eqnarray}
Due to the boundary condition $\delta A_0 = 0$, the BRST ghost must satisfy $\partial_0C = 0$ at the two boundaries in order to preserve $\delta A_0 = 0$. We also require $\bar{C}$ to satisfy $\partial_0 \bar{C} = 0$. The total bosonic one-loop fluctuation is (up to an overall factor of $1/4$)
\begin{eqnarray}
    (\delta A_i, \delta \phi, \phi^+_{\mu\nu}) (\partial_+\partial_- + \partial_\mu \partial^\mu)
    \begin{pmatrix}
        \delta A_i\\
        \delta \phi\\
        \delta \phi^+_{\mu\nu}
    \end{pmatrix}
    - \sum_{i=0}^5\bar{C}\partial_i\partial^i C
\end{eqnarray}
where $i$ ranges from $0$ to $5$ in the first term. Both $\delta \phi$ and $\delta A_1$ have Neumann boundary conditions, so their path integrals cancel the ghost path integral. We are left with the integrals of $A_0, A_\mu, \phi^+_{\mu\nu}$.

To compute the fermionic one-loop determinant, we perform a change of variables for the fermions \footnote{What we have done is to first replace $\bar{\lambda}\to -\bar{\lambda}$, take the sums and the differences of the fermions with and without overlines, and finally act on the fermions without overlines by $\gamma_5$ which reverses the sign in front of the fermionic vectors.}:
\begin{eqnarray}
    \begin{aligned}
        &\bar{\Psi} = 
    \begin{pmatrix}
        \bar{\psi}_\mu\\
        \bar{\psi}\\
        \bar{\psi}^+_{\mu\nu}
    \end{pmatrix} = 
    \frac{1}{2}
    \begin{pmatrix}
        -\lambda_\mu\\
        \lambda\\
        \lambda^+_{\mu\nu}
    \end{pmatrix} +
    \frac{1}{2}
    \begin{pmatrix}
        -\bar{\lambda}_\mu\\
        -\bar{\lambda}\\
        \bar{\lambda}^+_{\mu\nu}
    \end{pmatrix}\\
    &\Psi = 
    \begin{pmatrix}
        \psi_\mu\\
        \psi\\
        \psi^+_{\mu \nu}
    \end{pmatrix} =
    \frac{1}{2}
    \begin{pmatrix}
        \lambda_\mu\\
        \lambda\\
        \lambda^+_{\mu\nu}
    \end{pmatrix} -
    \frac{1}{2}
    \begin{pmatrix}
        \bar{\lambda}_\mu\\
        -\bar{\lambda}\\
        \bar{\lambda}^+_{\mu\nu}
    \end{pmatrix}
    \end{aligned}
\end{eqnarray}
The boundary conditions on the fermions translate to the following boundary conditions on the new variables:
\begin{eqnarray}
    \begin{aligned}
    &\partial_0 \psi_\mu = \psi = \psi^+_{\mu\nu} = 0\\
    &\bar{\psi}_\mu = \partial_0\bar{\psi} = \partial_0 \bar{\psi}^+_{\mu\nu} = 0
    \end{aligned}
\end{eqnarray}
The fermion kinetic terms become
\begin{eqnarray}
    \begin{aligned}
        &(\psi_\mu, \psi, \psi^+_{\mu\nu})i\partial_1
    \begin{pmatrix}
        \psi_\mu\\
        \psi\\
        \psi^+_{\mu\nu}
    \end{pmatrix}
    +
    (\bar{\psi}_\mu, \bar{\psi}, \bar{\psi}^+_{\mu\nu})i\partial_1
    \begin{pmatrix}
        \bar{\psi}_\mu\\
        \bar{\psi}\\
        \bar{\psi}^+_{\mu\nu}
    \end{pmatrix}\\
    &+ 2 (\psi_\mu, \psi, \psi^+_{\mu\nu})
    \begin{pmatrix}
        -\partial_0 & -d & -2d^\dagger \\
        d^\dagger & \partial_0  & 0\\
        \frac{1}{2}(d + *d) & 0 & \partial_0
    \end{pmatrix}
    \begin{pmatrix}
        \bar{\psi}_\mu\\
        \bar{\psi}\\
        \bar{\psi}^+_{\mu\nu}
    \end{pmatrix}
    \end{aligned}
\end{eqnarray}
where $d^\dagger$ is minus the divergence on the first index:
\begin{eqnarray}
    d^\dagger (\lambda_\mu) = -\nabla^\mu \lambda_\mu \quad d^\dagger(\lambda^+_{\mu\nu}) = -\nabla^\mu \lambda^+_{\mu\nu} 
\end{eqnarray}
The fermion kinetic term can be written in a compact form:
\begin{eqnarray}
    (\bar{\Psi}, \Psi)
    \begin{pmatrix}
        i\partial_1 & \slashed{D}_5 \\
        \slashed{D}_5 & i\partial_1
    \end{pmatrix}
    \begin{pmatrix}
        \bar{\Psi}\\
        \Psi
    \end{pmatrix}
\end{eqnarray}
where $\slashed{D}_5$ is the 5 dimensional Dirac operator
\begin{eqnarray}
    \slashed{D}_5 = 
    \begin{pmatrix}
        -\partial_0 & -d & -2d^\dagger \\
        d^\dagger & \partial_0  & 0\\
        \frac{1}{2}(d + *d) & 0 & \partial_0
    \end{pmatrix}
\end{eqnarray}
The mode matching is similar to that of the 2d $\sigma$-model in the previous chapter. Each bosonic mode $(\delta A_\mu, \delta A_0, \delta\phi^+_{\mu\nu})$ is matched to two fermionic modes
\begin{equation}
    \begin{pmatrix}
        \bar{\Psi} \\
        \Psi
    \end{pmatrix}
    \in
    \left\langle
    \begin{pmatrix}
        \delta A_\mu, \delta A_0, \delta\phi^+_{\mu\nu}\\
        0
    \end{pmatrix},
    \begin{pmatrix}
        0\\
        \slashed{D}_5(\delta A_\mu, \delta A_0, \delta\phi^+_{\mu\nu})
    \end{pmatrix}
    \right\rangle
\end{equation}
The determinant of the Dirac operator in this basis cancels the bosonic determinant  $\slashed{D}_5^2 = \partial_0\partial_0 + \partial_\mu \partial^\mu$. The unmatched modes are annihilated by $\slashed{D}_5$ and therefore annihilated by the 5d Laplacian $\partial_0\partial_0 + \partial_\mu \partial^\mu$. The boundary condition for $\delta A_0, \delta \phi^+_{\mu\nu}$ imply that they both vanish. The massless modes $\delta A_\mu$ are tangent vectors to the correspondence $M_{k, k + 1}$ \footnote{First we notice that massless modes $\delta A_\mu$ are time-independent away from the time slice containing the instanton operator. Next, the standand tangent-obstruction complex for abelian instantons,
\begin{eqnarray}
    0 \to \Omega_0(\C^2) \to \Omega_1(\C^2) \to \Omega_2^+(\C^2) \to 0
\end{eqnarray}
where the first map is the exterior derivative $d$ acting on scalars and the second map is $d$ composed with the projection onto self-dual two-forms, shows that the middle cohomology group is annihilated by $d$ and $d^\dagger$ and hence annihilated by the Laplacian.
} and can be expanded in Fourier modes along the $x^1$ circle similar to \eqref{eq: fourier}. As a result, the one-loop determinant at $(Y_k, Y_{k + 1})$ is
\begin{eqnarray}
    \frac{1}{\hat{a}(TM_{k, k+1}(Y_k, Y_{k + 1}))}
\end{eqnarray}
So the final result of the amplitude in the 6d gauge theory is
\begin{eqnarray}
    \bra{f} a_+^{6d} \ket{g} = \sum_{Y_k\subset Y_{k + 1}}\frac{f(Y_{k + 1})g(Y_k)}{ \hat{a}(TM_{k, k+n}(Y_k, Y_{k + 1}))}
\end{eqnarray}
A similar computation without the insertion of the instanton operator yields the norm formula \eqref{eq: overlap}.
Hence we have the desired result:
\begin{eqnarray}
    a_+ = a_+^{6d}
\end{eqnarray}
Therefore, $a_+$ defined in chapter \ref{ch: proof} is identified with the charge 1 instanton line operator in the 6d gauge theory. A similar argument shows that $a_-$ is identified with the charge $-1$ instanton line operator.
\acknowledgments
BZ would like to thank Samuel Crew for suggesting this project. We thank Ian Grojnowski, Zhipu Zhao, Xuanchun Lu, Amihay Hanany, Andrei Negut for helpful discussions. We would also like to thank Cyril Closset, Panos Betzios, Mathew Bullimore and Samuel Crew for helpful comments on the first draft. This work has been partially supported by STFC consolidated grant ST/X000664/1.

\appendix

\section{Young Diagrams}\label{app: young_diag}
\ytableausetup{boxsize=1em}
In this paper, we draw Young diagrams in the second quadrant. Here is an example:
\begin{eqnarray}\label{ydiagram_example}
  Y = \ydiagram{4,3,1}
\end{eqnarray}
A boxes in a Young diagram have coordinates $(p, q)$, where $p$ increases to the right and $q$ increases to the bottom. Here is an example with the coordinates labelled:
\begin{eqnarray}
  \ytableausetup{boxsize=2.5em}
  Y =
  \begin{ytableau}
    (0, 0) & (1, 0) & (2, 0) & (3, 0) \\
    (0, 1) & (1, 1) & (1, 2) \\
    (0, 2)
  \end{ytableau}
\end{eqnarray}
We always identify the box with coordinates $(p, q)$ with the monomial $q_1^pq_2^q$, where $q_1, q_2$ are the equivariant parameters acting on $U(1)$ instanton moduli spaces on $\C^2$. Here are the monomials associated with the boxes:
\begin{eqnarray}
  \ytableausetup{boxsize=2em}
  Y =
  \begin{ytableau}
    1 & q_1 & q_1^2 & q_1^3 \\
    q_2 & q_1q_2 & q_1^2q_2 \\
    q_2^2
  \end{ytableau}
\end{eqnarray}

For a Young diagram $Y_k$ of $k$ boxes, we denote the set of boxes that can be added to (resp. removed from $Y_k$) and still yield a valid Young diagram $Y_{k + 1}$ (resp. $Y_{k - 1}$) by $\operatorname{Add}(Y)$ (resp. $\operatorname{Rm}(Y)$). Since we identify a box $(p, q)$ with its weight $q_1^pq_2^q$, the addable and removable boxes in the diagram $Y = \ytableausetup{boxsize=0.5em}\ydiagram{4,3,1}$ can be written as:
\begin{align*}
  \ytableausetup{boxsize=1em}
\operatorname{Add}(Y) &= \left\{
  \ydiagram[*(green)]{1} \in
  \ydiagram{4,3,1}*[*(green)]{4+1}*[*(green)]{0,3+1}*[*(green)]{0,0,1+1}*[*(green)]{0,0,0,1}
  \right\} = \{ q_1^4, q_1^3 q_2, q_1 q_2^2, q_2^3\}\\
\operatorname{Rm}(Y) &=
\left\{
\ydiagram[*(red)]{1} \in
  \ydiagram{4,3,1}*[*(red)]{3+1}*[*(red)]{0,2+1}*[*(red)]{0,0,1}
  \right\} = \{ q_1^3, q_1^2 q_2, q_2^2\}.
\end{align*}

\bibliographystyle{JHEP}
\bibliography{Instanton_Operators}

\end{document}